\documentclass[11pt,superscriptaddress, titlepage]{revtex4-1}  
\usepackage{amsmath}  
\usepackage{graphicx}  
\usepackage{dcolumn}   
\usepackage{bm}        
\usepackage{amssymb}   
\usepackage[colorlinks,hyperindex]{hyperref}
\usepackage{hyperref}
\usepackage{mathptmx}
\usepackage[T1]{fontenc}
\usepackage{lmodern}
\usepackage{fancyhdr}
\fancypagestyle{report}%
\hypersetup
	{
		colorlinks,%
		citecolor=blue,%
		linkcolor=black,%
		urlcolor=black,%
	}

\renewcommand{\thesection}{\Roman{section}}
\begin{document}
\widetext
\title{\Large{Direct searches of Type III seesaw triplet fermions at high energy $e^+e^-$ collider}}
\author{Deepanjali Goswami}
\thanks{g.deepanjali@iitg.ernet.in}
\affiliation{Department of Physics, Indian Institute of Technology Guwahati, Assam 781039, India.}
\author{P. Poulose}
\thanks{poulose@iitg.ernet.in}
\affiliation{Department of Physics, Indian Institute of Technology Guwahati, Assam 781039, India.}

\begin{abstract}
The signatures of heavy fermionic triplets ($\Sigma$) arising in scenarios like Type III seesaw model are probed through their direct production and subsequent decay at high energy electron-positron collider.
Unlike the case of LHC, the production process has strong dependence on 
 the mixing parameter ($V_{e,\mu}$), making the leptonic collider unique to probe such mixing. We have established that with suitably chosen
kinematic cuts,  a 1 TeV $e^+e^-$ collider could probe the presence of $\Sigma$ of mass in the range of 500 GeV having $V_e=0.05$ with a few inverse femto barn luminosity through single production. The cross section
is found to be not sufficient to probe the case of triplet-muon mixing through single triplet production. On the other hand, the pair production considered at 2 TeV centre of mass energy is capable of probing both the mixing scenarios
efficiently. Studying the mass reach, presence of charged fermionic triplets upto a mass of about 980 GeV could be established at $3\sigma$ level through single production at a 1 TeV $e^+e^-$ collider with moderate
luminosity of 100 fb$^{-1}$, assuming  $V_e  = 0.05$ . The pair production case requires larger luminosity, as the cross section is  smaller in this case. With an integrated luminosity of
300  fb$^{-1}$, the mass reach in this case is close to 1 TeV with triplet-muon mixing, while it is slightly lower at about 950 GeV in the case of  $V_\mu  = 0.05$.

\end{abstract}


\maketitle
\newpage
\section{Introduction}
\label{intro}
\noindent 
The Standard Model (SM) of particle physics has established itself firmly as the description of dynamics of elementary particles at the electroweak scale. All measurements at the LHC conform to this, including the information on the Electroweak Symmetry Breaking (EWSB). However, many reasons including the lack of mechanism to generate masses for neutrinos, lack of candidate for dark matter, inability to explain the baryon asymmetry of the universe, along with other technical issues like the mechanism to stabilise the Higgs boson mass against quantum corrections, force us to look beyond the SM. It is expected that the new physics should show up in the TeV range of energies. Concerning the mechanism to generate mass to the neutrinos, the see-saw mechanism \cite{seesaw, Chikashige:1980ui}
 has emerged as the most popular and perhaps the most viable way of generating tiny mass of the observed light neutrinos of three different flavours. The seesaw mechanism  effectively exploits this idea by introducing a lepton number violating Majorana mass terms, either directly or generated dynamically. The tininess of the neutrino mass \cite{Ade:2013zuv} in this case is achieved with the help of large mass scale present in the scenario, usually brought in as the mass of a heavy partner. Generically, the seesaw mechanism is categorised into three types. In the Type I seesaw model \cite{seesaw}, a minimum of two gauge singlet right-handed neutrino fields are introduced in addition to the SM fields. In this case, the light neutrino mass is inversely proportional to the mass of this new partner fermion.  In Type II seesaw model \cite{typeII},  $SU(2)_{L}$ triplet scalar fields with hypercharge $Y=2$ are introduced, the vacuum expectation value (vev)  acquired by which induces Majorana mass to the neutrinos. In type III seesaw model \cite{typeIII}  fermionic triplet fields with $Y= 0$ are introduced, with a Yukawa term involving the SM lepton doublet and the SM Higgs field, and with Majorana mass terms. This third scenario leaves both charged as well as neutral heavy fermions in the spectrum, which could be searched for at the colliders. In principle, such fermions could be heavy, and out of reach of the LHC. At the same time, it is possible that such additional fermions have masses in the range of TeV, and thus could possibly be searched for at the LHC and at the proposed high energy leptonic colliders like the International Linear Collider (ILC) \cite{ILC, ILCpolarisation} or the Compact Linear Collider (CLIC) \cite{CLIC}.  We shall refer these high energy leptonic collider facilities as the Future Leptonic Colliders (FLC).
The phenomenology of Type-III seesaw model  in the context of LHC has been carried out in some detail by many authors 
\cite{  arXiv:hep-ph0703080, arXiv:0805.1613, delAguila:2008cj,Li:2009mw, Bandyopadhyay:2011aa,Eboli:2011ia, vonderPahlen:2016cbw, Ruiz:2015zca}. Experimental searches for the additional charged, as well as neutral heavy fermions arising in this model are performed by both CMS and ATLAS. Considering data at $\sqrt{s}=13$ TeV, CMS \cite{CMS:2016hmk}  has set a lower limit of 430 GeV on the triplet mass, whereas the ATLAS results \cite{ATLAS-CONF,Aad:2015cxa} ruled out masses in the range below 325 - 540 GeV under specific scenarios considered, with the larger value obtained with the assumption of decay exclusively to $W \ell$ of the  neutral fermion, and to $W\nu$ in the case of charged triplet. The Yukawa interaction term, that leads to the off-diagonal mass matrix for the neutral fermions, also causes mixing in the charged lepton sector. Simultaneous presence of the mixing with two flavours
receive very stringent constraints from  Lepton Flavour Violating (LFV) decays like $\mu \rightarrow e\gamma, ~ eee$, and $\tau \rightarrow \mu \gamma, ~ e\gamma$. However, if the mixing is restricted to single flavour, it could be large enough (constrained by the electroweak precision data)  to leave its effects at the colliders. The latter case is, but, not possible to probe at the LHC, when restricted to pair production of the heavy fermions, as the mixing parameter cancels out in the branching fraction. Cross-section for single production of heavy fermion in association with SM leptons has the potential to probe the mixing at the production level. However, this cross-section is too small to investigate at the LHC. 
On the other hand, at the leptonic colliders, the production itself is sensitive to the mixing, as we shall describe in details later.  Single production of the charged and neutral heavy fermions in the electon-proton collider (LHeC) is studied in Ref.~\cite{Liang:2010gm}.  While there are studies of indirect influence of the presence of  triplet fermions in the context of Higgs pair production at the ILC \cite{ILCpheno}, the direct production is not explored to the best of our knowledge.
The advantages of the leptonic colliders, being sensitive to the mixing at the production level, as well as their clean environment, are exploited in the present study in which we shall investigate the possible reach of high energy $e^-e^+$ collider in searching for heavy fermions, and discuss the sensitivity to the mixing. We may note that  although the study is made in the context of the Type III seesaw model, the conclusions can be  easily adapted to any model in which such triplet fermions are present. 

We focus our attention on the production of both charged as well as neutral fermion triplets at the FLC and explore the identification of these triplets over the SM backgrounds in different channels.  In particular, we shall discuss how  the mixing can be probed through the processes studied here. We may note that, in a realistic seesaw model we need at least two triplet fields in order to accommodate the observed mass splittings of the three neutrino flavours. However, in this study, for simplicity,  we shall consider a single family of triplet fermion field in addition to the SM fields. In a more realistic case, this may be considered equivalent to the case when the other fermions are much heavier, and therefore not relevant to the phenomenology at the energies considered.

We organise this article as follows. In Section \ref{typeiii} we shall discuss some details of the Type-III seesaw model. In Section \ref{process} we shall describe the processes under study, and discuss the results. Finally, we shall summarize and conclude in Section  \ref{conclusion}.

\section{Type-III seesaw model}
\label{typeiii}

In this section we shall describe the features of the Type III seesaw model relevant to our discussion. We have used the FeynRules implementation of the model as explained in the reference \cite{Biggio:2011ja}. Therefore,  for convenience, we shall follow the notations and conventions used in this reference. The Lagrangian involving the 
 $SU(2)_L$ triplet fermion field, denoted here as $\Sigma$, along with the SM part denoted by $\mathcal{L}_{SM}$ is given by
$
\mathcal{L} = \mathcal{L}_{SM} + \mathcal{L}_{\Sigma},\nonumber
$
with 
\begin{eqnarray}
\mathcal{L}_{\Sigma}&=&
\textrm{Tr\ensuremath{\left(\bar{\Sigma\,}\slash\!\!\!\!\!\!D\Sigma\right)}}-
\frac{1}{2}M_\Sigma~\textrm{Tr} \left(\overline{\Sigma}\Sigma^{c}+\bar{\Sigma}^{c}\Sigma\right)
- \sqrt{2}Y_{\Sigma l}~\left(\tilde{\phi}^{\dagger}\bar{\Sigma}L
- \bar{L}\Sigma\tilde{\phi}\right)
\label{eq:lagrangian}
\end{eqnarray}
\noindent where $\mathit{M}_{\Sigma}$ is the mass parameter of the triplet
and $\mathit{Y}$$_{\Sigma l}$ is the Yukawa couplings corresponding to the lepton flavours $l=e,~\mu,~\tau$.   The left-handed lepton doublets of the SM is denoted by $\mathit{L}\equiv(\nu, l)^{T}$, and the Higgs doublet by $\phi$$\equiv$($\phi^{+},$$\phi^{0})^{T}$$\equiv$($\phi^{+}$,($\mathit{v}$+$\mathit{H}+$ $\mathit{i}$$\eta$)/$\sqrt{2}$ )$^{T}$, with  $\tilde{\phi}$ = $\mathit{i}$ $\tau_{2}$ $\phi^{*}$. The fermion triplet $\Sigma$ is explicitly given by
\begin{equation}
\Sigma = \begin{pmatrix}\Sigma^{0}/\sqrt{2} & \Sigma^{+}\\
\Sigma^{-} & -\Sigma^{0}/\sqrt{2}
\end{pmatrix}
\end{equation}
and its conjugate is denoted by $\Sigma^{c}\equiv\mathit{C}\bar{\Sigma}^{T}$, where $C$ is the charge conjugation operator.

The two-component charged spinors are combined into Dirac spinor \(
\Psi \equiv \Sigma^{+c}_{R} + \Sigma^{-}_{R}\), with  \(
\Psi_R \equiv \Sigma^-_{R}\), and  \(
\Psi_L \equiv \Sigma^{+c}_{R},\)
to conveniently express the mixing of the SM charged leptons with the triplets, whereas the neutral component, $\Sigma^0$ is left as the two-component Majorana fermion. The  Lagrangian in the new set up is given by

\begin{eqnarray}
\mathcal{L}_{\Sigma} & = & \overline{\Psi}i \slash\!\!\!\!\!{\partial}\Psi + \overline{\Sigma}^{0}_{R}i \slash\!\!\!\!\!{\partial}\Sigma^{0}_{R}- gW^{3}_{\mu}\overline{\Psi}\gamma^{\mu} \Psi + g\left( W^{+}_{\mu} \overline{\Sigma}^{0}_{R}\gamma^{\mu} P_{R} \Psi + W^{+}_{\mu}\overline{\Sigma}^{0c}_{R} \gamma^{\mu}P_{L}\Psi + h.c.\right)- \overline{\Psi} M_{\Sigma}\Psi-  \nonumber \\&& 
\left(\frac{1}{2}\overline{\Sigma}^{0}_{R}~M_{\Sigma}~\Sigma^{0c}_{R} + h.c \right)  
 - \left(\phi^{0}\overline{\Sigma}^{0}_{R}Y_{\Sigma}\nu_{L} + \sqrt{2}\phi^{0}\overline{\Psi}Y_{\Sigma}\ell_{L} + \phi^{+}\overline{\Sigma}^{0}_{R}Y_{\Sigma}\ell_{L} -      \sqrt{2}\phi^{+}\overline{\nu}^{c}_{L}Y^{T}_{\Sigma}\Psi + h.c. \right) 
\end{eqnarray}
\label{lagrangian}

In Appendix A1, we provide the expanded form of the Lagrangian in the mass basis. Mixing between the heavy fermion, $\Sigma$ and the SM leptons are denoted by $V_{\alpha} = \frac{v}{\sqrt{2} M_{\Sigma}} ~Y_{\Sigma {\alpha}}$, with $\alpha=e,~ \mu,~\tau$. When two of these parameters are present simultaneously, they are bound by experimental measurements from the flavour changing rare decays $\mu \rightarrow e\gamma$, $\tau \rightarrow \mu\gamma$ and $\tau \rightarrow e\gamma$ given by \cite{typeIIImodel, Abada:2007ux, delAguila:2008pw, Biggio:2011ja}
\begin{eqnarray}
&&\vert V_{e} V_{\mu} \vert < 1.7 \cdot 10^{-7},~~
\vert V_{e}V_{\tau} \vert < 4.2 \cdot 10^{-4},~~
\vert V_{\mu}V_{\tau} \vert < 4.9 \cdot 10^{-4}
\end{eqnarray}
Single parameter bounds are obtained from Electroweak Precision Measurements \cite{delAguila:2008pw}, with the present bounds given by
\begin{eqnarray}
&&\vert V_{e} \vert < 0.055,~~ 
\vert V_{\mu} \vert < 0.063, ~~
\vert V_{\tau} \vert < 0.63 
\end{eqnarray}
The off-diagonal charged-current and neutral current interactions allow the triplet fermions to decay to the SM final states involving leptons, gauge bosons and the Higgs boson. The decay widths of different channels are given by \cite{Franceschini:2008pz}

\begin{eqnarray}
\Gamma(\Sigma^{0} \rightarrow l^{-}_{\alpha}W^{+}) &=& \Gamma(\Sigma^{0} \rightarrow l_{\alpha}^{+}W^{-}) = \frac{g^{2}}{64 \pi} \vert V_{\alpha} \vert^{2} \frac{M^{3}_{\Sigma}}{M^{2}_{W}} \Big( 1- \frac{M^{2}_{W}}{M^{2}_{\Sigma}}\Big)^{2}  \Big(1+ 2\frac{M^{2}_{W}}{M^{2}_{\Sigma}}\Big)\nonumber \\ 
\sum\limits_{l} \Gamma(\Sigma^{0} \rightarrow \nu_{l}Z) &=& \frac{g^{2}}{64\pi c^{2}_{W}} \sum\limits_{\alpha} \vert V_{\alpha} \vert^{2} \frac{M_{\Sigma}^{3}}{M_{Z}^{2}} \Big(1 - \frac{M^{2}_{Z}}{M^{2}_{\Sigma}}\Big)^{2} \Big(1 + 2\frac{M^{2}_{Z}}{M^{2}_{\Sigma}}\Big)\nonumber\\
\sum\limits_{l} \Gamma(\Sigma^{0}\rightarrow \nu_{l}H)& =& \frac{g^{2}}{64\pi} \sum\limits_{\alpha}\vert V_{\alpha} \vert^{2} \frac{M^{3}_{\Sigma}}{M^{2}_{W}}\Big(1 - \frac{M_{H}^{2}}{M^{2}_{\Sigma}}\Big)^{2}\nonumber\\
\hspace{2cm}\sum\limits_{l} \Gamma(\Sigma^{+} \rightarrow \nu_{l}W^{+}) &=& \frac{g^{2}}{32\pi} \sum\limits_{\alpha} \vert V_{\alpha} \vert^{2} \frac{M_{\Sigma}^{3}}{M^{2}_{W}} \Big(1 - \frac{M^{2}_{W}}{M^{2}_{\Sigma}} \Big)^{2} \Big(1 + 2\frac{M^{2}_{W}}{M^{2}_{\Sigma}}\Big) \nonumber\\
\Gamma(\Sigma^{+} \rightarrow l^{+}_{\alpha}Z) &=& \frac{g^{2}}{64\pi c^{2}_{W}} 
\vert V_{\alpha} \vert^{2} \frac{M^{3}_{\Sigma}}{M^{2}_{Z}} \Big(1 - \frac{M^{2}_{Z}}{M^{2}_{\Sigma}}\Big)^{2} \Big(1 + 2\frac{M^{2}_{Z}}{M^{2}_{\Sigma}}\Big)\nonumber\\
\Gamma(\Sigma^{+} \rightarrow l^{+}_{\alpha}H) &=& \frac{g^{2}}{64\pi} \vert V_{\alpha} \vert^{2}\frac{M_{\Sigma}^{3}}{M^{2}_{W}}\Big(1 - \frac{M^{2}_{H}}{M^{2}_{\Sigma}}\Big)^{2}
\label{eq:decays}
\end{eqnarray}
\vskip 5mm

As mentioned in the Introduction, the LHC bounds on the mass of the heavy fermions is slightly below 500 GeV. For our study, we shall consider $M_\Sigma=500$ GeV. However, we shall present the dependence of the results and conclusions on $M_\Sigma$ in the discussions that follow. 
The decay branching ratios (BR) of the triplets to channels specified in Eq.~\ref{eq:decays} are given in Table~\ref{table:BR} for $M_\Sigma=500$ GeV.    Note that the BR is independent of $V_\alpha$, when only one such mixing is present.  Of the charged (neutral) triplets, about 51 percent decay to $W\nu~(W\ell)$, and 26 percent to $Z\ell~(Z\nu)$, with  23 percent decaying to $H\ell~(H\nu)$.  These fractions remain the same for masses above 500 GeV.

\begin{table}[h]
\begin{center}
\small
\begin{tabular}{|l|l|c|}
\hline
 Decay $\Sigma^\pm$ & Decay $\Sigma^0$ & BR in $\%$ \\ \hline \hline 
$\Sigma^{\pm} \rightarrow W^{\pm}\nu$ & $\Sigma^{0} \rightarrow W \ell$ & 51\\ \hline
$\Sigma^{\pm} \rightarrow Z\ell^{\pm}$ & $\Sigma^{0} \rightarrow Z\nu$ & 26\\\hline
$\Sigma^{\pm} \rightarrow H\ell^{\pm}$ & $\Sigma^{0} \rightarrow H\nu$ & 23 \\ \hline
\end{tabular}
\caption{Branching ratio of the charged and neutral triplet fermion with mass, $M_\Sigma=500$ GeV.  }
\label{table:BR}
\end{center}
\end{table}

The production mechanism being largely independent of the mixing,  it is hard to obtain information regarding mixing parameters at LHC. Firstly, the pair production mechanisms involve gauge couplings of the triplets, and therefore the dependence on mixing is not significant.
 The decay widths, on the other hand have strong dependence on the mixings. However, in the total cross section, which is a  product of  production cross section and branching ratio of the decay channel considered, this dependence is cancelled, as long as the heavy flavour mixes with one flavour of the SM leptons. The advantage of FLC in this regard is evident, as the production mechanism itself could depend on the electron-triplet mixing parametrised by $V_e$,  directly through the couplings of the form $e \Sigma V$, where $V= W, Z$ . 

\section{Direct Production of the Triplets  }
\label{process}
We shall consider the single as well as pair production of both the neutral and  charged triplet fermions at the high energy versions of the ILC. 

\subsection{Single production of $\Sigma^{0}$ and $\Sigma^\pm$}

\begin{figure}[h]
\includegraphics[width = 5cm]{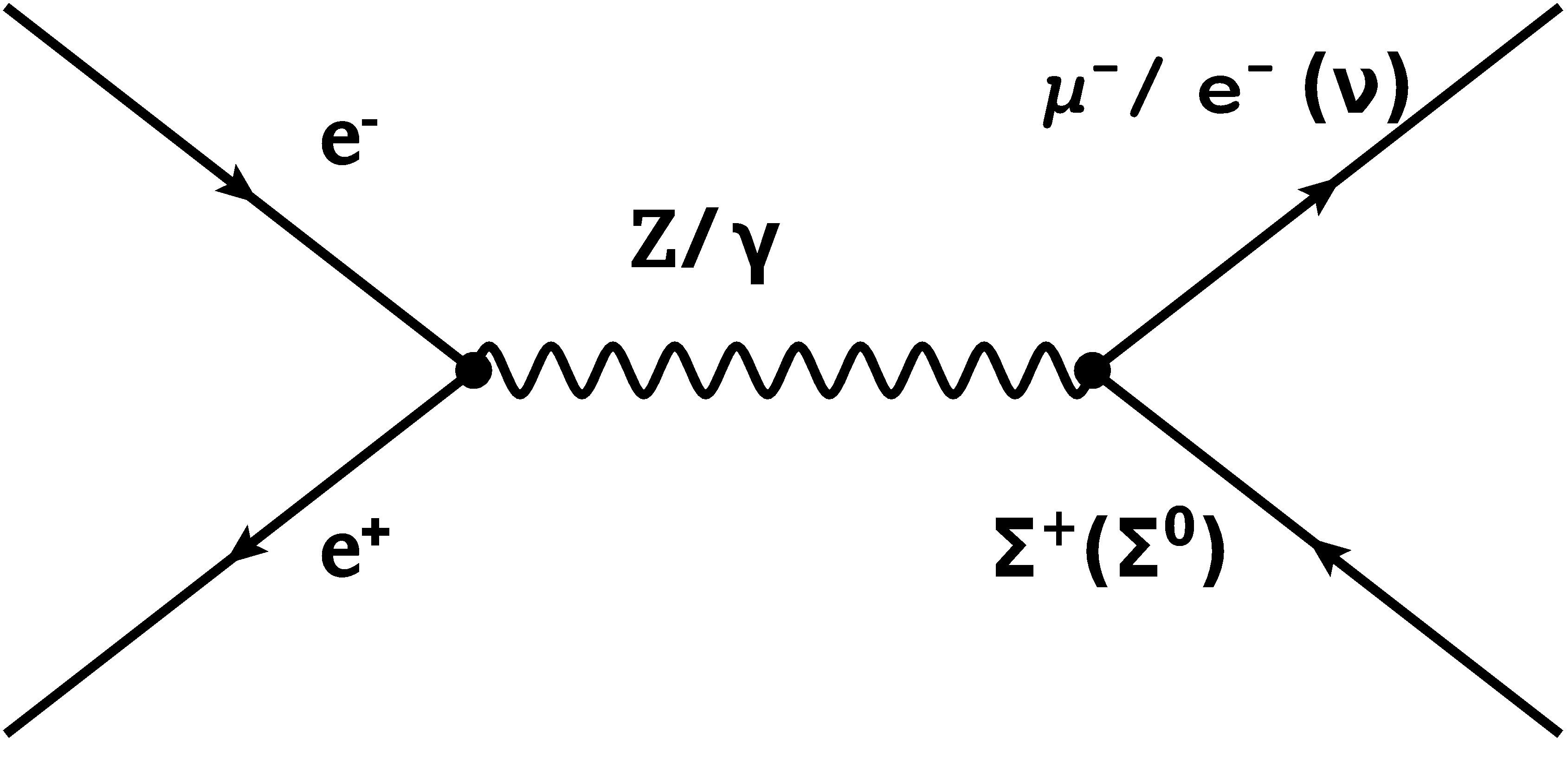}\hskip 10mm
\includegraphics[width=3cm]{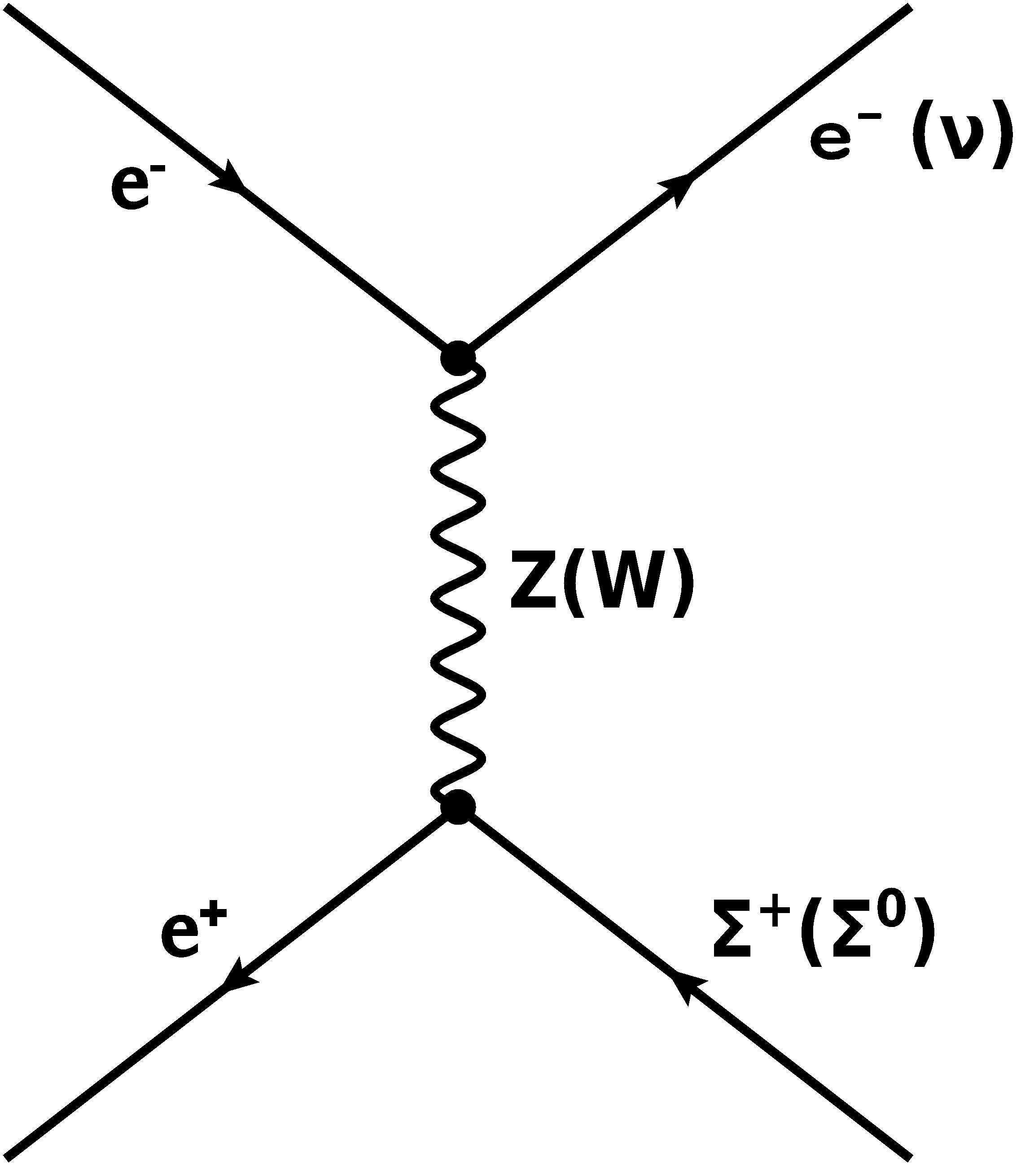}
\caption{Feynman diagrams contributing to the process $e^{+} e^{-} \rightarrow \Sigma^{+} \ell^{-}$ ($\Sigma^{0}\nu$).}
\label{fig:fd-single}
\end{figure}

The single production of neutral and charged components of the fermion triplet along with a neutrino or lepton, respectively, are sensitive  to the mixing of these heavy fermions with the SM leptons at the production level. The Feynman diagrams involve an $s$-channel exchange of gauge bosons. In addition, when $V_e\ne0$, the process receive a $t$-channel contribution, as shown in Fig.~\ref{fig:fd-single}. 

\begin{figure}[h]
\includegraphics[width=10cm]{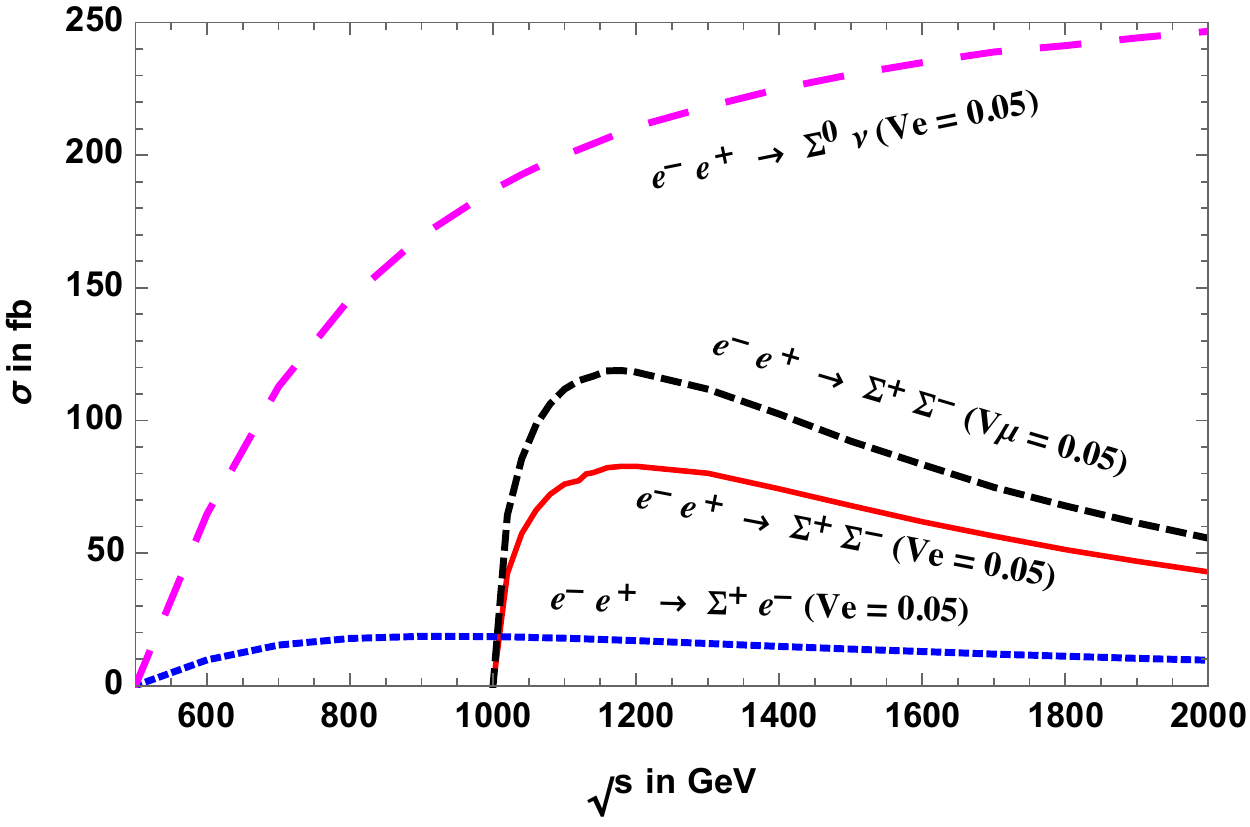}
\caption{Cross section for $e^+e^-\rightarrow \Sigma^0\nu,~~\Sigma^\pm e^\mp,~~\Sigma^+\Sigma^-$ against the centre of mass energy, with $M_\Sigma=500$ GeV. }
\label{fig:csection}
\end{figure}

Complying with the direct limits discussed in the Introduction, we shall consider $M_\Sigma=500$ GeV  for our numerical studies.
The expressions for cross sections of different cases are given in Appendix A2. From the Feynman diagrams, it is clear that the cross section is proportional to  $\sum\limits_\alpha |V_\alpha|^2$ in case of neutral triplet production, and to the individual $|V_\alpha|^2$ in the case of charged triplet production. Note that $\Sigma^0\nu$ production does not have a photon mediated $s-$channel contribution. It is expected that the $s$-channel contribution falls down with $\sqrt{s}$, and thus become negligible at high energies considered here. 
On the other hand,  the $t$-channel contribution and the interference between the $t$- and the $s$-channel give substantial contribution when $V_e\ne 0$. We consider two different cases of (i) $V_e\ne 0,~~V_\mu=0$, and (ii) $V_e= 0,~~V_\mu\ne 0$, with $V_\tau=0$ in both cases. The first case leads to $e^+e^-\rightarrow e^\mp\Sigma^\pm, ~~\Sigma^0\nu$ through both the $s$- and $t$-channels, whereas the second case leads to $e^+e^-\rightarrow \mu^\mp\Sigma^\pm, ~~\Sigma^0\nu$ through purely $s$-channel process.  Cross sections for the latter case is very small, and we shall not consider this in our further analysis. Figure \ref{fig:csection} shows the cross section against the centre of mass energy, with the cross section for $\Sigma^0\nu$ grows to a saturation of 250 fb at around 2 TeV, while $e^\mp\Sigma^\pm$ production cross section saturates at 18 fb around 1 TeV centre of mass energy. 
We fix our centre of mass energy at 1 TeV, where the cross section for neutral single production is sizeable, with 187 fb. 

The heavy fermions further decay as per Eq.~\ref{eq:decays}, leading to $W\ell \nu,~Z\nu\nu$ and $H\nu\nu$ final states in the case of $\Sigma^0$, and $W\ell \nu,~Z\ell\ell$ and $H\ell\ell$ final states in the case of $\Sigma^\pm$ productions.  With the further decay of $W,~Z$ and $H$, this leads to the detector level final states of $2j+ E\!\!\!\!\!\slash,~~2b+E\!\!\!\!\!\slash$ (arising only from $\Sigma^0$ production), $2b+2\ell,~~2j+2\ell,~~2\ell^++2\ell^-$  (arising only from $\Sigma^\pm$ production), and $2j+\ell+E\!\!\!\!\!\slash$ and $2\ell+E\!\!\!\!\!\slash$ (arising from both $\Sigma^0$ and $\Sigma^\pm$ productions).  The cross sections corresponding to these final states, along with the SM backgrounds are given in Table~\ref{table:finalstate-single}.  The cross sections quoted are the fiducial cross sections including the respective branching fractions obtained from Madgraph \cite{madgraph} with basic generation level cuts on the transverse momenta of the jets and leptons, $p_T(j)\le 20$ GeV, $p_T(l) \le 10$ GeV, and pseudorapidity of $|\eta| \le 2.5$ employed. 
The $2j+E\!\!\!\!\!\slash$ coming from neutral triplet has large continuum QCD background. Similarly, the purely leptonic channel, $2e^+2e^-$, and channels with $\tau\bar{\tau}$ coming from the Higgs bosons have small cross section. We have therefore focused on the other cases of purely leptonic and semi-leptonic final states, as well as the $b\bar b+E\!\!\!\!\!\slash$, where the $b-$ quark pair arises from the $H$ decay. 

\begin{table}[ht!]
\begin{center}
\small
\begin{tabular}{|c||c||c||c|}
\hline \hline
Final State & Process ($e^+e^- \rightarrow  \Sigma^\pm e^- , ~~~\Sigma^0 \nu$~) & \multicolumn{2}{c|}{$\sigma\times$ {BR} in fb} \\\cline{3-4}
&&Signal& Background\\
\hline
\hline
2$j$ + $e^{-}$+ E\!\!\!\!\! \slash & $\Sigma^{+} e^{-}$ $ \rightarrow W^+ e^- ~\nu$ &  32.7 &  WWZ(0.5),  WW(74.5), \\\cline{2-2}
& $\Sigma^{0} \nu$ $ \rightarrow W^{+} e^{-} \nu$&  & $t\tilde{t}(1.68)$, ZZ(2.17), $Zjj$(2.77) \\\hline \hline
 2$j$ + $e^{-}$ $ e^{+}$ & $(\Sigma^{+} e^{-}+\Sigma^{-} e^{+})$ $ \rightarrow Z e^{+} e^{-}$ &  4.2  &   $e e$jj(34.5)\\\hline \hline
$e^{-}$ $e^{+}$ + E\!\!\!\!\! \slash & $(\Sigma^{+} e^{-}+\Sigma^{-} e^{+}) \rightarrow W^{\pm} e^{\mp} \nu,~~Z e^{-} e^{+}$ & 14.8 &   WW(14.09),  $WWZ$(0.036) \\\cline{2-2}
& $\Sigma^{0} \nu$ $ \rightarrow W^{\pm} e^{\mp} \nu,~~Z \nu \nu$ & &  ZZ(0.35), $t\tilde{t}(1.6)$  \\\hline \hline
2$e^{-} +2 e^{+}$ & $(\Sigma^{+} e^{-}+\Sigma^{-} e^{+}) \rightarrow $ Z $e^{\pm} e^{\mp}$ & 0.3 & ZZ(0.065), $eeee$(3.6) \\\hline \hline
{$b \bar{b}$ $e^{+} e^{-}$} & {$(\Sigma^{+} e^{-}+\Sigma^{-} e^{+}) \rightarrow H ~ e^{\pm} e^{\mp}$} & {7.2} &  {HZ(0.27)}, {ZZ(0.78)}\\\hline \hline
{$b \bar{b} + E\!\!\!\!\! \slash$} & {$\Sigma^{0} \nu \rightarrow H ~\nu \nu$} & {37.6} & {HZ(2.1 ), ZZ(8.9)}\\\hline \hline
{2j + E\!\!\!\!\! \slash} & {$\Sigma^{0} \nu \rightarrow Z ~\nu \nu$} & {22.3} & {$q \bar{q}$(440.1)}\\\hline \hline
$\tau^{+} \tau^{-} + E\!\!\!\!\! \slash $ & $\Sigma^{0} \nu \rightarrow H \nu \nu$ & 1.6  & HZ(0.09), ZZ(1.05)\\ \hline  \hline
$\tau^{+}\tau^{-}~ e^{+} e^{-}$ &   {$(\Sigma^{+} e^{-}+\Sigma^{-} e^{+}) \rightarrow H ~ e^{\pm} e^{\mp}$}  & 0.16  & HZ (0.0125)\\ \hline \hline
\end{tabular}
\caption{Signal and corresponding background fiducial cross sections corresponding to the different final states arising from the process $e^{-} e^{+} \rightarrow \Sigma^{\pm} e^{\mp}$ and $e^{+} e^{-} \rightarrow \Sigma^{0} \nu$, with $p_T(j)\ge 20$ GeV, $p_T(l) \ge 10$ GeV, and pseudo rapidity $|\eta| \le 2.5$ for jets and leptons. Centre of mass energy of $\sqrt{s}$ = 1000 GeV and $M_{\Sigma}$ = 500 GeV are considered, along with the assumed mixing of $V_{e}=0.05,~~V_\mu=V_\tau=0$.}
\label{table:finalstate-single}
\end{center}
\end{table}

To analyse these selected final states, we generated 50000 events in each case using Madgraph5 with the in-built Pythia6 \cite{Sjostrand:2006za} used for ISR, FSR, and showering and hadronization. The basic generation level cuts are those quoted above, with $p_T(j)\ge 20$ GeV, $p_T(l) \ge 10$ GeV, and $|\eta| \le 2.5$ for the jets as well as leptons. 
These events are then passed on to Madanalysis5\cite{Conte:2012fm} to analyse and optimise the final selection criteria. Fastjet \cite{Cacciari:2011ma} is used for jet reconstruction with anti-$k_T$ algorithm and jet radius of $R=0.4$. For the detector simulation, Delphes3  \cite{Ovyn:2009tx, deFavereau:2013fsa} with standard ILD card is used. Before applying any selection cuts, proximity check for leptons are done with leptons closer than $\Delta R_{jl}=0.4$ are ignored. Further selection was based on the required number of final state leptons and jets, and considering the distinguishability of the kinematic distributions. In Table~\ref{table:finalstate-single-significance} the cut-flow chart is presented along with the final significance that is expected at an integrated luminosity of 100 fb$^{-1}$. 
We shall briefly discuss the selection cuts of each of the final states below.

\begin{enumerate}
\item \underline{$2j+e^-+E\!\!\!\!\! \slash$}

 The signal and background events after the basic generation level cuts are 3273 and 8170, respectively. After demanding that the event should contain two jets and one electron, and veto-ing the presence of $b$-jet, the number of events reduce to 2187 and 3871 for the signal and background, respectively. The $b$-jet veto is used to reduce the $t\bar t$ background events.  This is followed by the selection of events with $ 100 ~ $GeV $\le p(e^-)$ ,  $p(j_1) \le 300$ GeV and $p(j_2) \le 200$ GeV  which reduces about 7\% of the background events, at the same time keeping about 77\% of the signal events. This leaves 285 background events against a  signal of 1681.  Overall, about 51\% of the original signal events are retained, against about 3.5\% of the background events. 
 
 Assuming only statistical uncertainty, signal significance computed with formula, $\frac{S}{\sqrt{S+B}}$, where $S$ is the number of signal events and $B$ is the number of background events, is 37.9 at the luminosity of 100 fb$^{-1}$ considered. In order to accommodate the systematic uncertainties, we have considered the following formula,
 \begin{equation}
 S_{\rm sys} = \frac{S}{\sqrt{S+ B+\alpha^2~B^2+\beta^2~S^2}},
 \label{eq:significance}
 \end{equation}
 where $\alpha$ and $\beta$ are the systematic uncertainties in the background and signal events, respectively. Systematics at leptonic colliders like ILC are expected to be well under control. Assuming a very conservative value of 5\% uncertainty in both the signal and background cases, we obtain a significance of 17.5 at the integrated luminosity of 100 fb$^{-1}$. 

 \item \underline{$2j+e^-e^+$}
 
 In this case, $p(e^-) \ge 140$ GeV and $p(e^+) \ge 140$ GeV,  and a selection of invariant mass of electron-positron pair, $M_{e^+e^-} > 200$ GeV, apart from demanding that there be one electron and one positron, and two jets are employed  to reduce the background from 3450 to its 27.5$\%$, while retaining 64.3 \% of the signal events. Now the background is further reduced to 110 by cut on the pseudo rapidity of the leptons,  $\eta (e^+) < 1$ and  $\eta (e^-) > -1$ . This selection leaves the signal events mostly unaffected.  A signal significance of 13.8 and 11 without and with assumed systematics could be achieved through this selection.

 \item  \underline{ $e^-e^++E\!\!\!\!\! \slash$}
 
 Here, electron positron pairs are more back to back compared to those in the  signal events. Demanding lepton separation, $\Delta R (e^+ , e^-) < 4$ reduces the background to  479 from 1620, while  keeping 1014 signal events  starting from 1489 events.
 This leads to a signal significance of about 26.2 without any systematics, which goes down to 14.8 with the assumed systematic uncertainties.
 
 \item  \underline{$b\bar b+e^{-}+e^{+}$}
 
A cut on the invariant mass of the lepton pair, $M_{e^+ e^-}>140$ GeV, apart from demanding two $b$-jets, one electron and one positron, takes away all the backgrounds, leaving 180 signal events with signal significance of 13.4 without systematics uncertainty and 11.1 with systematic uncertainty.

\item  \underline{$b\bar b +E\!\!\!\!\! \slash$}

In this case,   $\Delta R (b, \bar{b}) > 0.6$ reduces the  background events  from 1100 to its 140, while the signal is reduced from 3760 to 1194. The corresponding signal significance without systematics is 32.6, which is reduced to 16.9 with the assumed systematics.
\end{enumerate}

\begin{table}[h!]
\begin{center}
\small
\begin{tabular}{|c||c||c|c||c||c|}
\hline \hline
Final State &  Selection cuts  &\multicolumn{2}{c|}{No. of events} & ~~$\frac{S}{\sqrt{S+B}}$~~&   $S_{\rm sys}$ \\\cline{3-6}
&  (All figures, except $N$ are  in GeV) &Signal & Backgd & \multicolumn{2}{c|}{ $V_{e}=0.05, V_{\mu}=V_{\tau}=0 $}    \\ \hline \hline
$2j+e^{-}+E\!\!\!\!\!\slash$ & No cut & 3273 & 8170 & & \\
& N(j) =2 ,N$(e^{-})$ = 1 ,N(b) = 0  & 2187 & 3871  &  & \\
&  $p(e^{-}) >  $ 100,  $p(j_1) < $ 300, $p(j_2) < $ 200  &  1681 & 285 & 37.9 &  17.5 \\ \hline \hline
 $2j + e^{-}e^{+}$ & No cut & 420 & 3450 & & \\
& N($e^{+}$) = 1, N$(e^{-})$ = 1, N(j) = 2 &  273 & 1500 & & \\
 & $p(e^-),~~ p( e^+) >$ 140,  M($e^{+} e^{-}) >$  200 & 270 & 948 &  &  \\
  & $\eta(e^+) < $ 1, $\eta(e^-) >  -1$  & 269 & 110 &  13.8 & 11.0  \\
 \hline \hline
$e^{-} e^{+} + E\!\!\!\!\!\slash$ & No cut & 1489 & 1620  & & \\
& N($e^{\pm}$) = 1, N(b) = 0 &  1103 & 1036 & &  \\
&  $\Delta R (e^+, e^-) <$ 4  &  1014 &  479 & 26.2 &  14.8 \\\hline \hline
$b \bar{b}+e^{+}e^{-}$ & No cut & 718  &  105 & &  \\
&N$(e^{+})$ = 1,  N($e^{-}$) = 1, N(b) = 2 & 180  & 14  & &  \\
& M($e^{+} e^{-}) >$ 140   & 180 &0  & 13.4 & 11.1 \\\hline \hline
$b \bar{b} + E\!\!\!\!\!\slash $ & No cut & 3760 & 1100 & & \\
&N($e^{+}$) $=$ 0, N($e^{-}$) = 0, N(b) = 2  &  1243 & 221  &  & \\
& $\Delta R (b, \bar{b}) >$ 0.6   & 1194 & 140 & 32.6 &  16.9 \\\hline \hline
\end{tabular}
\caption{ The cut-flow and signal significance for different final states arising from the single production of $\Sigma^{0}$ and $\Sigma^{\pm}$ at  $\sqrt{s}$ = 1000 GeV and 100$fb^{-1}$ luminosity for processes $e^{-} e^{+} \rightarrow \Sigma^{\pm} \ell^{\mp}$ and $e^{+} e^{-} \rightarrow \Sigma^{0} \nu$,  with  
$M_{\Sigma}$ = 500 GeV. }
\label{table:finalstate-single-significance}
\end{center}
\end{table}

Assuming that the kinematics of both the background and signal events remain more or less the same, we can scale the luminosity to the required value for signal significance of $5\sigma$. In Table~\ref{table:lumi_sig5} we present the projected requirement of luminosity for this case, along with the expected number of signal and background events after the selection criteria adopted as in Table~\ref{table:finalstate-single-significance}.  The $2j+ e^- + E\!\!\!\!\!\! \slash$ final state gives the best case scenario with  about 2 fb$^{-1}$ luminosity leading to $5\sigma$ sensitivity and  the purely leptonic channel of $e^+e^-+E\!\!\!\!\! \slash$ requires about 4 fb$^{-1}$. 

\begin{table}[h!]
\begin{center}
\small
\begin{tabular}{c||c||c||c}
\hline \hline
Final state&$\int{\cal L}$ (in fb$^{-1}$)&$S$&$B$\\ 
&for $\frac{S}{\sqrt{S+B}} = 5\sigma$&&\\ \hline \hline
$2j + e^{-}+ E\!\!\!\!\! \slash$& 2 & 33.6 &   5.7  \\ \hline \hline
$b \bar{b} + E\!\!\!\!\! \slash $& 2.5 &  29.8 &   3.5 \\ \hline \hline
$e^{-} e^{+} + E\!\!\!\!\! \slash$& 4 &  40.5 & 19.1  \\  \hline \hline
$b \bar{b}+e^{+}e^{-}$ & 14 & 25.2  & 0   \\    \hline \hline
$2j + e^{-}e^{+}$ & 14 & 37.6 & 15.4   \\  \hline
 \hline
\end{tabular}
\caption{ Luminosity requirement for signal significance  of 5$\sigma$ for different final states of the processes $e^{-} e^{+} \rightarrow \Sigma^{\pm} \ell^{\mp}$ and $e^{+} e^{-} \rightarrow \Sigma^{0} \nu$ at  $\sqrt{s}$ = 1000 GeV 
with  $M_{\Sigma}$ = 500 GeV, for the case of $V_e=0.05,~~V_\mu=V_\tau=0$, along with the signal $S$ and background $B$ at the specified luminosities.}
\label{table:lumi_sig5}
\end{center}
\end{table}

\subsection{Pair production of $\Sigma$}
\begin{figure}[h]
\includegraphics[height=2.9cm]{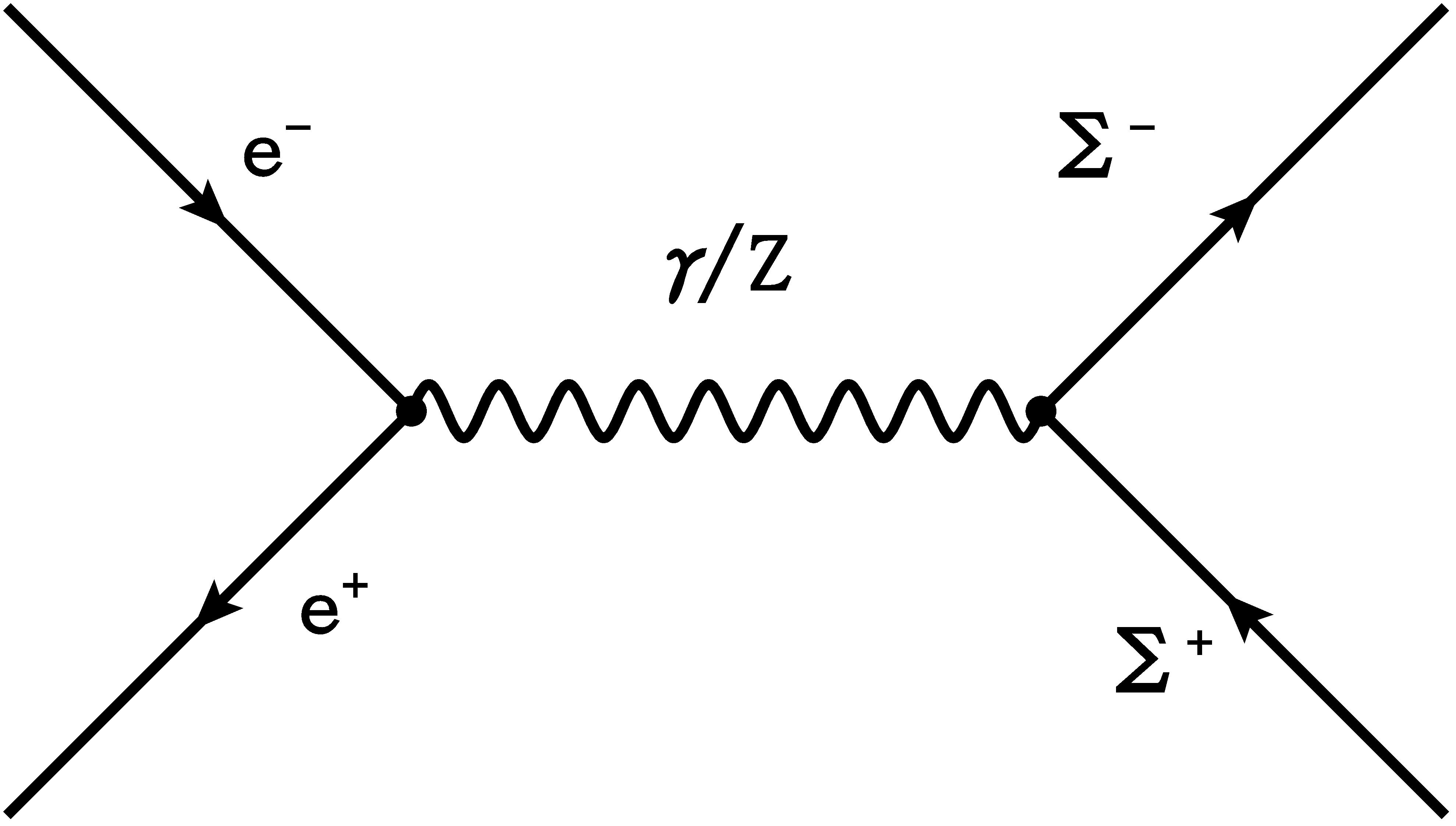} \hskip 10mm
\includegraphics[height=3.6cm]{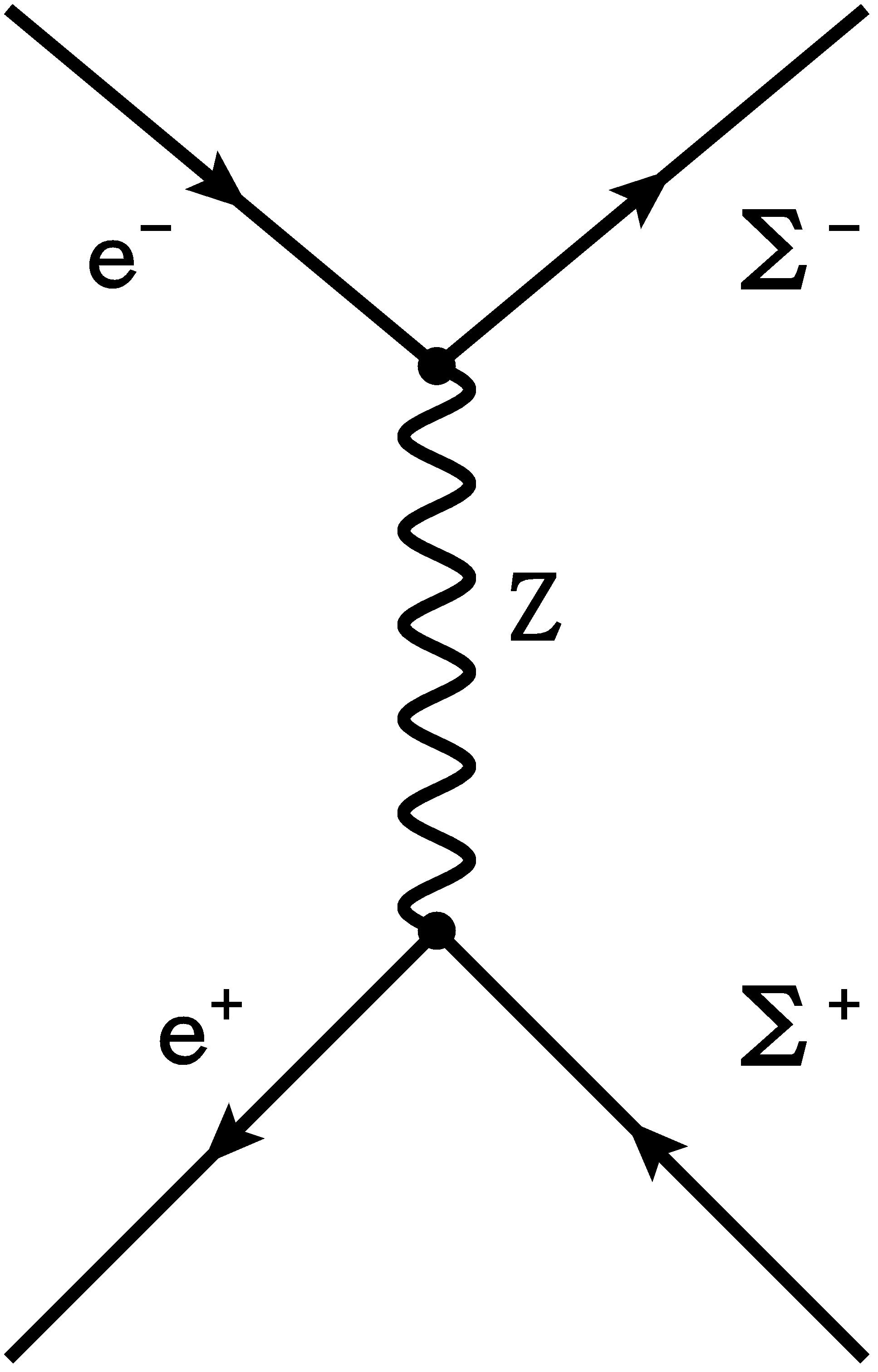}
\caption{Feynman diagrams contributing to the process $\mathit{e}^{+}e^{-}\rightarrow$ $\Sigma^{+}\Sigma^{-}$ in type-III seesaw model.}
\label{fig:fd-pair}
\end{figure}

We shall next consider the pair production of the  triplet fermions. The Feynman diagrams corresponding to the production of charged fermion pairs are shown in Fig.\ref{fig:fd-pair}. The neutral fermion pair production also goes through the same channels, except the one with the photon exchange.
Notice that  the $t$-channel contribution to the cross section here is proportional to the fourth power of the mixing parameter $V_\ell$. Thus, it is expected that the $s$-channel dominates. Again, the $s$-channel for $\Sigma^0$ pair production is proportional to the square of the $Z\Sigma^0\Sigma^0$ vertex, which is proportional to the $|V_\ell|^2$. Thus, the cross section for neutral fermion pair production is very small. On the other hand, the $Z\Sigma^+\Sigma^-$ vertex is proportional to $\left(|V_\ell|^2 - 2~\cos^2 \theta_W\right)-|V_\ell|^2~\gamma^5$, and therefore receives a sizeable contribution even in the absence of mixing. In addition, the $V_e\ne 0$ case has a $t$-channel contribution, which is similar to that of the case of $\Sigma^0$ pair production. This results in a small difference between the two cases of $V_e\ne 0 $ and $V_\mu\ne 0$, with the former slightly smaller than the latter, indicating destructive interference between the $s$- and the $t$-channel processes. We reiterate that this advantage of the FLC, where the production is sensitive to the mixing is absent at the LHC.  In the following, we shall consider only the pair production of the charged fermions. The cross section against the centre of mass energy is given in Fig.~\ref{fig:csection}. At $\sqrt{s}=2$ TeV, the cross section is 43 fb and 55.7 fb corresponding to the cases of $V_e = 0.05$ and $V_\mu=0.05$, respectively, with the other two mixings taken to be absent. The values corresponding to the peak of the cross section at about 1.2 TeV centre of mass energy with values of 83 fb and 119 fb, respectively for the above two cases.

With the decay of $\Sigma^\pm$ to $W\nu,~~Z\ell$ or $H\ell$, and the subsequent decays of $W, ~~Z$ and $H$ considered, we have the purely hadronic final states of $4j+E\!\!\!\!\! \slash$, semi-leptonic final states of $4j+2\ell,~4j+\ell+E\!\!\!\!\! \slash,~2j+4\ell,~2j+3\ell+E\!\!\!\!\! \slash,~2j+2\ell+E\!\!\!\!\! \slash,~2j+\ell+E\!\!\!\!\! \slash, ~4b+2\ell$ and the purely leptonic case of $2\ell+E\!\!\!\!\! \slash$. We have included only the case of Higgs decay to $b$-pair, as the other cases come with much smaller effective cross section. Again, $Z$ to $b$-pair decay is not included. In Table~\ref{table:ch-finalstates} the cross sections of these final states arising from the signal for the two cases of (i) $V_e=0.05,~V_\mu=V_\tau=0$, and (ii) $V_\mu=0.05,~V_e=V_\tau=0$ are given, along with the corresponding SM background cross sections. The cross sections are obtained from the MC simulation with Madgraph5 with Pythia6 used for hadronisation and showering. We have included the generation level basic cuts on the transverse momenta of jets and leptons of $p_T(j)>20$ GeV and $p_T(\ell)>10$ GeV, and considered jets and leptons with pseudo rapidity of $|\eta|<2.5$. The final states with $2j+4\ell$ and $2j+3\ell+E\!\!\!\!\! \slash$  have very small cross sections, and therefore require luminosities at the level of inverse femtobarn to probe these channels. The  purely leptonic final state of $2\ell+E\!\!\!\!\! \slash$ comes with large SM background of  about three orders larger than the signal. In our further analysis we do not consider these three cases. 

\begin{table}
\begin{center}
\small
\begin{tabular}{|c||c||c|c||c|}
\hline
\hline
Final State & Process & \multicolumn{3}{c|}{$\sigma\times$ {BR} in fb} \\\cline{3-5}
&($e^+e^- \rightarrow \Sigma^+\Sigma^-$) &\multicolumn{2}{c|}{Signal}&\multicolumn{1}{c|}{Background}\\ \cline{3-4}
&&$V_e=0.05$&$V_\mu=0.05$&\\
\hline
\hline
$4j+E\!\!\!\!\! \slash$ &$W^+ W^- \nu\nu$ & 1.3 & 2.0 & $WWZ$ (1.4),  $WW \nu \nu (16.6)$\\\hline \hline
$4j+ \ell \ell $ & $ZZ \ell^+ \ell^- $ & 0.3 & 0.4 &  $WWZ(0.15)$, $ZZjj(0.7)$ \\\hline \hline
$4j+\ell +E\!\!\!\!\! \slash$ & $W^+ Z \ell^- \nu$ & 0.6 & 0.8 & $t\tilde{t}(1.4)$, $WWjj(0.5)$ \\\hline \hline
$2j+4\ell$ & $ZZ \ell^+ \ell^- $ & 0.02 & 0.03 &  $ZZZ(0.0002)$ \\\hline \hline
$2j+3\ell+E\!\!\!\!\! \slash$ & $W^+ Z \ell^- \nu$ & 0.04  & 0.05  & $WWZ(0.03)$\\\hline \hline
$2j+2\ell+E\!\!\!\!\! \slash$ & $ZZ\ell^+ \ell^- ,~~W^+ Z \ell^- \nu$ & 0.4 & 0.5 & WWZ(0.12),$t\tilde{t}(0.44)$\\\hline \hline
$2j+\ell+E\!\!\!\!\! \slash$ & $W^+ W^- \nu\nu,~~W^+ Z \ell^- \nu$ & 0.8 & 1.0 & $WWZ$(0.27),$WW$(12.2), t$\tilde{t}$(1.4), $ZZ(0.1)$\\ \hline \hline
$2\ell+E\!\!\!\!\! \slash$& $W^+ W^- \nu\nu$ & 0.1 & 0.2 &  $WW$(3.4), $t \bar{t}$(0.43), $\ell\ell\nu\nu$(181.7)\\\hline \hline
$2b2\bar{b}+2\ell$ & $HH \ell^+ \ell^- $ & 1.7 & 2.2 & $HHZ(0.004)$\\\hline \hline
\end{tabular}
\caption{Final state fiducial cross sections of the signal from $e^{+}e^{-} \rightarrow \Sigma^{-}\Sigma^{+}$, and the corresponding SM background processes, with the selection of $p_T(\ell) \ge 10$ GeV , pseudo rapidity of leptons $|\eta_{\ell}| \le 2.5$ and the  selection of $p_T(j)\ge 20$ GeV, $|\eta_{j}| \le 2.5$. Centre of mass energy of $\sqrt{s}$ = 2 TeV, and $M_{\Sigma}$ = 500 GeV are considered. The lepton in the final state $\ell$ is $e$ or $\mu$ for the cases of $V_e=0.05$ and $V_\mu=0.05$, respectively. }
\label{table:ch-finalstates}
\end{center}
\end{table}

\begin{table}[h!]
\small
\begin{tabular}{|c||c||c|c|c|c||c|c|c|c|}
\hline
 &Selection cuts  & \multicolumn{4}{c||}{$V_{e} =0.05$} & \multicolumn{4}{c|}{$V_{\mu} = 0.05$} \\\cline{3-10}
Final State & (All dimensional quantities & signal&backd &$\frac{S}{\sqrt{S+B}}$&$S_{\rm sys}$&signal & backd &$\frac{S}{\sqrt{S+B}}$&$S_{\rm sys}$ \\
&  are in GeV)& $S$ &  $B$ & && $S$ & $B$ & &\\\hline\hline
{\small $4j+E\!\!\!\!\! \slash$} & $N(j) = 4$ , $p(j_{1})>100$  &147 & 1679 & 3.4  & 1.5 & 243 & 2164 & 5 & 2 \\\cline{2-10} 
&$N(j) \ge 3$, $p(j_{1})>100$    & 353 & 3914 & 5.4 & 1.7   & 503 & 3914 & 7.5 & 2.4  \\\hline \hline
 &  $N(\ell^{\pm}) = 1$, $N(j) = 4$,   $ N(b) = 0$  &  50 & 12  & 6.3 & 6 & 73 & 13 & 7.8 & 7.3  \\ 
$4j+\ell^{\pm}+E\!\!\!\!\! \slash$  & $p(\ell^\pm ) >$ 100, $MET>100$ &&&&&&&&\\\cline{2-10}
& $N(\ell^{\pm}) = 1$, $N(j) \ge 3$,  $N(b) = 0$ &  &  & & &&&  & \\
& $p(\ell^\pm) > 100 $, $MET>100$ & 106 & 33 &  8.9  & 8.1 & 154& 39 & 11.1 & 9.6  \\\hline \hline
 &   $N(\ell^{\pm})=1$, $N(j) = 4$ , &  &    &   & && & &  \\
{ $4j+  \ell^+ \ell^- $} & $p(\ell^\pm)>100$, $\Delta R(\ell^{+}, \ell^{-}) \ge 2$  & 29 &  0 & 5.3 &  5.2 & 74 & 0 & 8.6 &  7.9 \\\cline{2-10}
 &$N(\ell^{\pm})=1$, $N(j) \ge 3$, &  &    &   & && & &  \\
& $p(\ell^\pm)>100$, $\Delta R(\ell^{+}, \ell^{-}) \ge 2$  & 56 & 0 & 7.4 &  7 & 140 & 0 & 11.8 & 10.1 \\\hline \hline
&  $N(\ell^{+})=1$, $N(\ell^{-}) = 1$,  $N(j) = 2$,   &  &  &&& &   &  & \\
{$2j+\ell^+ \ell^- +E\!\!\!\!\! \slash$}  &$N(b) = 0$, $p(\ell^-) >   100$  &  47 & 12 &  6.1 & 5.8  & 54 &  15 & 6.5 & 6.1\\\hline \hline
  & $N(\ell^{\pm}) = 1$, $N(j) = 2$ , $N(b) = 0$,  &  & &&  & &  &  & \\
{$2j+\ell^{\pm}+E\!\!\!\!\! \slash$} & $|\eta( \ell)| < 1$, $E(\ell) < 900$  & 87 & 365 & 4.0 & 3.0 & 121 & 10 & 10.5 & 9.3 \\
 & $p(j_{1}) < 600$, $p(j_{2})  < 300$  &  & & & & & & &\\\hline \hline
 & $N(\ell^{+}) = 1$, $N(\ell^{-}) = 1$, & 24 & 0 & 4.9 & 4.7 & 34 & 0  &  5.8 & 5.5  \\ 
{$4b+\ell^+ \ell^-$ }   & $N(b) = 4$ , $p(e^\pm) > 60 $ & & & & & & & & \\\cline{2-10}
 &  $N(\ell^{+}) = 1$, $N(\ell^{-}) = 1$,  & 114 & 0 & 10.6 & 9.4 & 163 & 0 & 12.7 &10.8\\\
  &$N(b) \ge  3$ ,  $p(e^\pm) > 60 $ & & & & & & & & \\\hline  \hline
\end{tabular}
\caption{ Number of surviving events,  and signal significance for different final states arising from the pair production of $\Sigma^{\pm}\Sigma^\mp$ at  300$fb^{-1}$ luminosity at  $\sqrt{s}$ = 2 TeV, and $M_{\Sigma}$ = 500 GeV at ILC.   $S_{\rm sys}$ corresponds to the signal significance with assumed systematics according to Eq.~\ref {eq:significance}. }
\label{table:finalstate-pair-significance}
\end{table}

The events generated are then passed on to Madanalysis5, using Fastjet for jet reconstruction with anti-$k_T$ algorithm and jet radius of $R=0.4$. Detector simulation was carried out with the help of Delphes3 with standard ILD card. Before applying any selection cuts, proximity check for leptons were done with leptons closer than $\Delta R_{jl}=0.4$ to the jets ignored. Further selection was based on the required number of final state leptons and jets, and considering the distinguishability of the kinematic distributions. In the $4j$ events, we considered two different  situations with (i) setting the number of jets exactly equal to four, and (ii) demanding every event has three jets or more. The second case provided with marginal improvement in the significance, and about double the signal events in each case.
In Table~\ref{table:finalstate-pair-significance} the cut-flow chart is presented along with the final significance that is expected at an integrated luminosity of 300 fb$^{-1}$. 
We shall briefly discuss the cuts used to optimise the selection below. 

\begin{enumerate}
\item \underline{$4j+E\!\!\!\!\! \slash$}\\[2mm]
With $p(j_1)>100$ GeV, the two cases of $N(j)=4$ and $N(j)\ge 3$ give significance of 3.4 and 5.4, respectively, for the scenario with $V_e\ne 0,~~V_\mu=V_\tau=0$ when only statistical errors are assumed. This is reduced to 1.5 and 1.7, respectively, with the assumed systematics of 5\% on both the signal and background event determination. The scenario with $V_\mu\ne 0,~~V_e=V_\tau=0$ has the corresponding significances of 5 (2) and 7.5 (2.4) considering statistical (statistical plus systematic) uncertainty. Notice that this channel is purely hadronic, and does not leave  any trace of the type of mixing involved.
\item \underline{$4j+\ell^\pm+E\!\!\!\!\! \slash$}\\[2mm]
Here $\ell$ is electron or muon depending on the case of $V_e\ne 0$ or $V_\mu\ne 0$. Unlike the case of  $4j+MET$, here the missing energy has a different topology in signal compared to that of the background (refer to Table~\ref{table:ch-finalstates} for the list of major backgrounds). A cut of  $p(\ell)>100$ GeV and $MET>100$ GeV are used apart from demanding one lepton and $N(j)=4$ or $N(j)\ge 3$, along with demanding $N(b)=0$ to reduce the $t\bar t$ background. The significance for the case of electron are 6.3 (6) and 8.9 (8.1) without (with) systematics assumed, for the two cases of jet counting of $(i) N(j)=4$ and $(ii) N(j)\ge 3$, respectively. In the case of muon, these are 7.9 (7.3) and 11.1 (9.6), respectively. Notice that the systematics have less pronounced effect here, as the events are small in number. We assume the charge of the lepton is identified, with both the cases giving similar results. 

\item \underline{$4j+\ell^+ \ell^-$}\\[2mm]
In this final state, the oppositely charged dileptons originate at the production in signal, whereas they come from the decay of $Z$ bosons in the case of the backgrounds. Therefore, the leptons are expected to be more energetic in the case of signal events. We employ a cut of $p(\ell^\pm)>100$ GeV  in both the cases of $N(j)=4$ and $N(j)\ge 3$ . In addition, we have assumed that the two leptons are separated with $\Delta R \ge 2$, as they are expected to be well separated in the case of signal events, whereas in the case of background events they will be more collimated as they originate from the $Z$ boson in flight. With these selection cuts, the background is practically eliminated. The significance for the four and three jet-counting are 5.3 and 7.5 for electrons, and 8.6 and 11.8 respectively for the case of muons. As the events are not very large, the systematics do not have much effect here. 
\item \underline{$2j+2\ell+E\!\!\!\!\! \slash$}\\[2mm]
Coming to the $2j+\ell+MET$ events, $p(\ell^-)>100$ GeV  is employed, leading to a significance of 6.1 and 6.5 for the case of electron and muon, respectively. Here again, the systematics have only a small role to play.

\item \underline{$2j+\ell^\pm+E\!\!\!\!\! \slash$}\\[2mm]
The major background here is the $WW$ production with the semi-leptonic decay of the pair. The lepton coming from the $W$ is expected to be very energetic, unlike the case of the signal. A cut on the energy of the lepton, $E(\ell)<900$ GeV is employed, along with a cut on the pseudo rapidity of lepton $|\eta(\ell)|<1$, reduced the background considerably. Further cuts on the momenta on jets $p(j_1) < 600$ GeV and $p(j_2) < 300$ GeV are considered to reach an expected significance of 4 (3) for electron without (with) systematics considered.
The case of muons presents a much better scenario with expected significance of 10.6 (9.3).

\item \underline{$4b+2\ell$}\\[2mm]
The background for the $4b$ events is quite suppressed. We have considered identifying two oppositely charged leptons, and the cases of $N(b)=4$ and $N(b)\ge 3$, along with demanding $p(e^{\pm}) > 60 $ GeV. The number of events surviving in the case of electron mixing are 24 and 114 respectively, with vanishing backgrounds in both cases. In the case of muon mixing, the significance is improved with the surviving number of events of 34 and 163, respectively. 
\end{enumerate}

\begin{figure}
\centering
\includegraphics[width = 8.15cm]{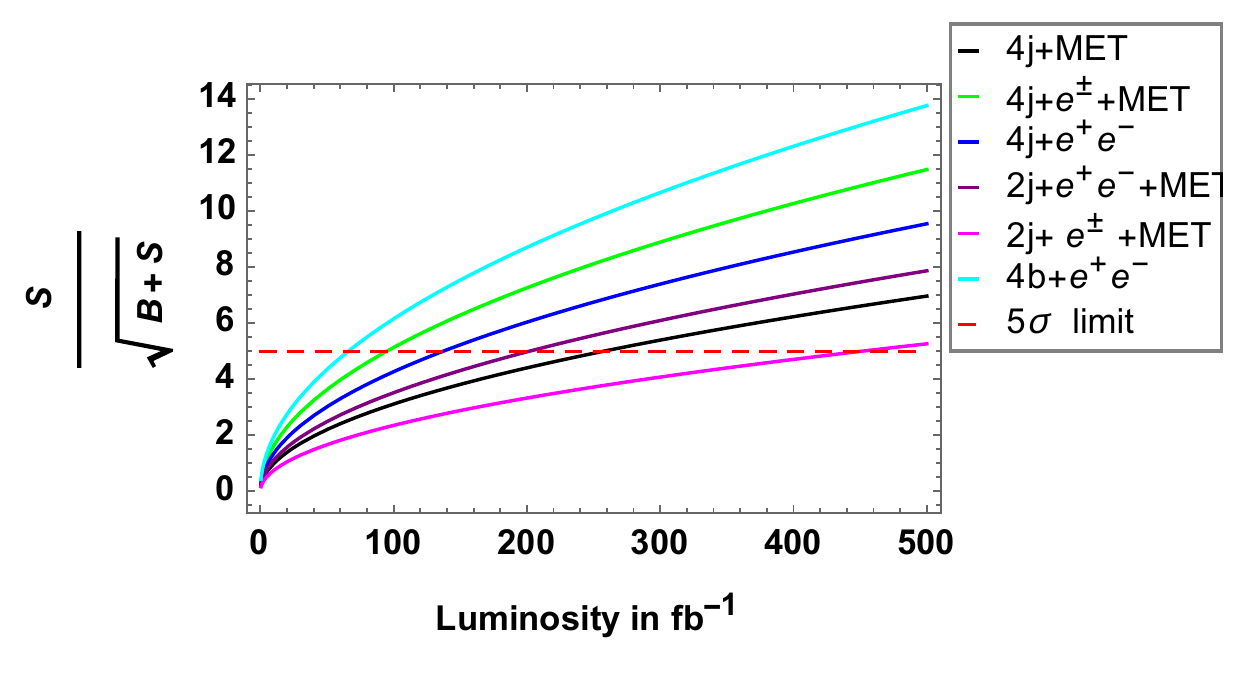}
\includegraphics[width = 8.15 cm]{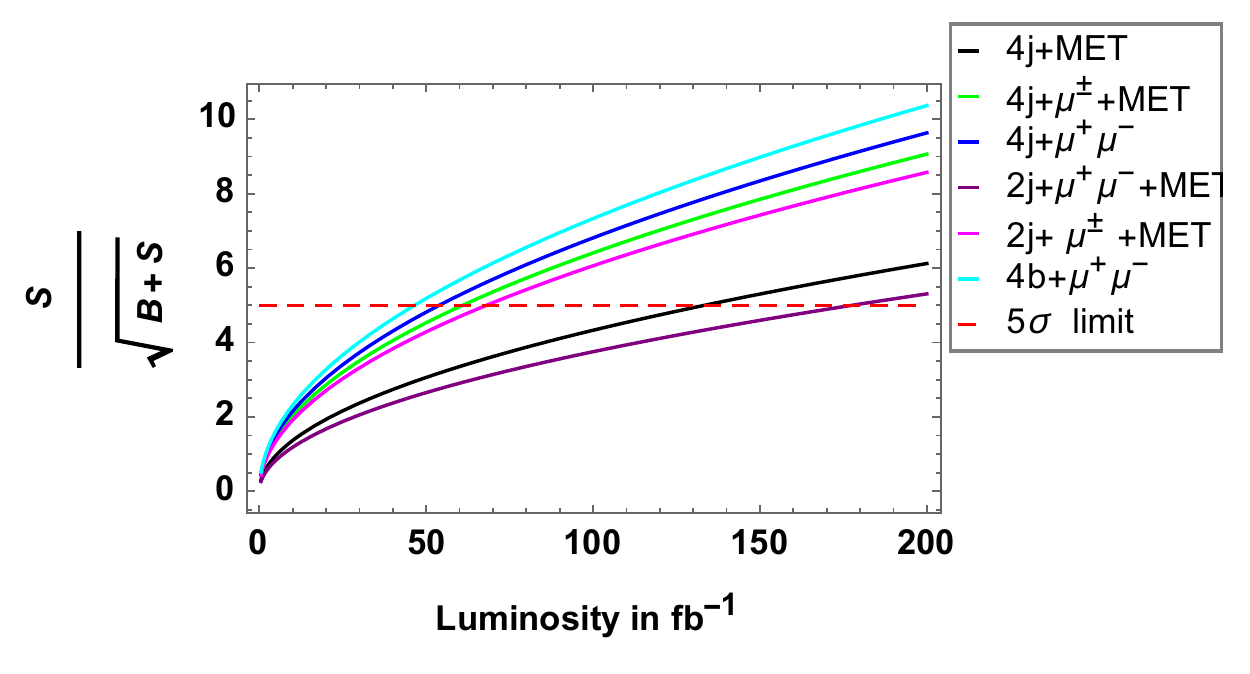}
\caption{Signal significance of different final states from pair production of $\Sigma^{\pm}$ against integrated luminosity at $\sqrt{s}=2$ TeV are shown. Mass of triplet fermion, $M_\Sigma=500$ GeV, and mixing parameters of $V_{e}=0.05$ (left) and $V_{\mu}=0.05$ (right) are considered with other mixings set to zero.     } 
\label{pairprodlum}
\end{figure}

Summarising, $4j+2\ell$ and $4b+2\ell$ provides the best case scenarios, where practically no background is present. Both of these cases could also distinguish the mixing scenarios from the flavour of the leptons produced. The single lepton events with missing energy accompanied by either four jets or two jets also provide very promising scenarios. Here the four jet case can distinguish the two mixing scenarios with the flavour identification, whereas the two jet case has the leptons arising also from the $W$ decay, and therefore, it will give a mixed signal.

 All the final states in both the cases with two different mixing scenarios  are used to indicate the projected luminosity required for $5\sigma$ significance in Fig.~\ref{pairprodlum}. Luminosity of less than 300 $fb^{-1}$ is sufficient to probe all the channels with electron in the final state (except $2j + e^{\pm} +E\!\!\!\!\! \slash$) at 5$\sigma$ level . On the other hand   $4j + E\!\!\!\!\! \slash$ and $2j+\mu^+ \mu^- +E\!\!\!\!\! \slash$ require about 130 and 180 $fb^{-1}$ luminosity, whereas all other channels with $\mu$ in the final state can be explored at $5\sigma$ level with less than 100 $fb^{-1}$ luminosity.

\subsection{Dependence on the mass of $\Sigma$}

\begin{figure}
\centering
\includegraphics[width = 13 cm]{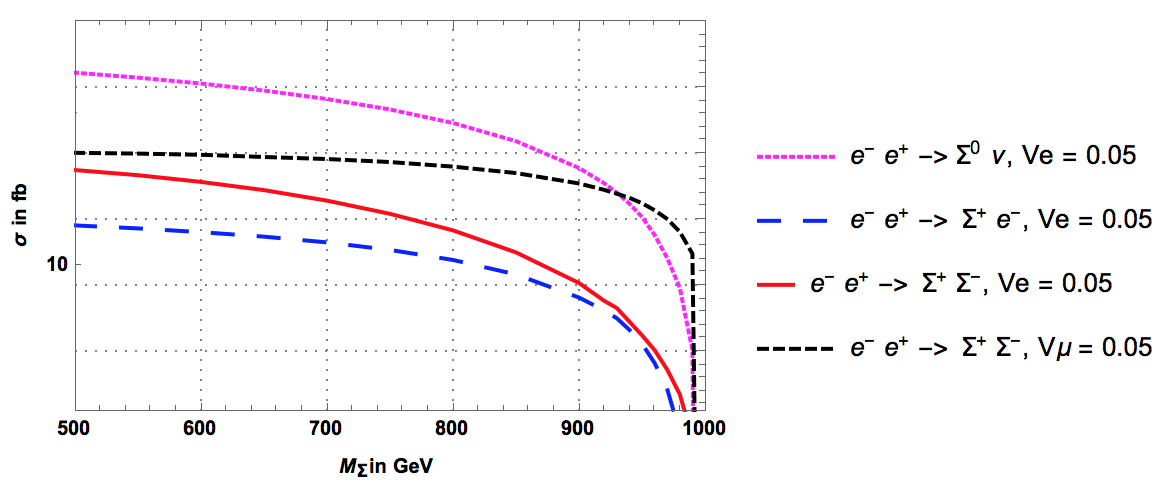}
\caption{Variation of cross-section with different mass of fermions for pair and single production at ILC. Single production is considered at $\sqrt{s}=1$ TeV, whereas $\sqrt{s}=2$ TeV is considered for pair production process. \label{figure:sig-mass}}
\end{figure}

In the analysis considered so far we had fixed the mass of the heavy fermion to $M_\Sigma=500$ GeV. In this section we shall briefly consider the mass dependence, and try to find an estimate of the reach of $M_\Sigma$ with the mixing fixed at $V_\ell=0.05$. Firstly, we plot the cross section
against $M_\Sigma$ in Fig.~\ref{figure:sig-mass}. The centre of mass energy considered for single production is 1 TeV, and that for pair production is 2 TeV. The near threshold behaviour of the pair production with $V_\mu\ne 0$ is distinctly different from the case with $V_e\ne 0$. This may be attributed to the fact that while the former case is a purely $s$-channel process, the latter has a contribution from the $t$-channel as well, facilitated by the presence of $Z\Sigma e$ coupling. The single production cases are presented only for $V_e\ne 0$ case, as the $\Sigma e$ production is not possible with $V_e=0$, while $\Sigma \nu$ production is very small in the case of $V_\mu\ne 0$. The mass dependence seems to follow the same pattern in the two cases of neutral as well as the charged fermion single production considered here. 
We shall now demonstrate that with 300 fb$^{-1}$ integrated luminosity, the reach of ILC is close to $M_\Sigma=1$ TeV. 
Let us consider the case of $4b+2\ell$ final state in the $\Sigma^+\Sigma^-$ pair production. The production cross section at $\sqrt{s}=2$ TeV is 43 fb with $V_e\ne 0$. The number of signal events left after the selection cuts is 114. This corresponds to a cross section times branching ratio of 0.38 fb. The selection cuts have eliminated the background, and thus number of signal events required for $3\sigma$ signal significance is about 9. This corresponds to a cross section times branching ratio of $\frac{9}{300}=0.03$. Assuming that the selection cuts behave the same way, the production cross section required to get this significance is $\frac{43}{0.38}\times 0.03=3.39$ fb. At $\sqrt{s}=2$ TeV, keeping $V_e=0.05$, but keeping all other parameters the same as the SM case, this cross section corresponds to a mass of $M_\Sigma=950$ GeV. A similar study of the $4j+2\ell$ and $4j+\ell+E\!\!\!\!\! \slash$ final states show that about $3\sigma$ significance is reached with a pair production cross section of 6.9 and 9.1 fb, respectively. These correspond to mass reaches of about  $910$ and 885 GeV, respectively.  Considering the $\mu$ channels with $V_\mu=0.05$, the situation get some what better with the addition of  $2j+\ell+E\!\!\!\!\! \slash$ also able to probe the model with $M_\Sigma$ very close to the kinematic limit of 1000 GeV. Table~\ref{massreach} summarises the mass reach at a 2 TeV ILC with an integrated luminosity of 300 fb$^{-1}$. 

\begin{table}[h!]
\small
\begin{tabular}{|c||c||c||c||c||c||c||c||c|}
\hline \hline
&\multicolumn{4}{c||}{$V_e=0.05$}&\multicolumn{4}{c|}{$V_\mu=0.05$}\\ \cline{2-9}
Final State &$S$&$B$& $\sigma(\Sigma^+\Sigma^-)$&$M_\Sigma$& $S$&$B$& $\sigma(\Sigma^+\Sigma^-)$&$M_\Sigma$\\
&&&in fb&in GeV&&&in fb&in GeV\\ \hline \hline
$4b+2\ell$& 9 & 0 & 3.4 & 945 & 9 & 0 & 3.0 & 997  \\ \hline \hline
$4j+2\ell$& 9 & 0 & 6.9 & 910 & 9.2 & 0.17 & 3.6 & 995 \\ \hline \hline
$4j+\ell^\pm+E\!\!\!\!\! \slash$& 16 & 33.3 & 9.1 & 885 & 23.7 & 38.5  & 8.5 & 990  \\ \hline \hline
$2j+\ell+E\!\!\!\!\! \slash$& 62 & 365 & 30.7 & 660  & 15 & 9.7  & 6.9 & 992 \\ \hline 
\hline
\end{tabular}
\caption{ The Mass reach at 2 TeV with integrated luminosity of 300 fb$^{-1}$ from selected channels of pair production of charged fermions, giving $3\sigma$ sensitivity. The corresponding production cross sections $\sigma(\Sigma^+\Sigma^-)$, and the number of signal ($S$) and background ($B$) events after the selection cuts are also given.}
\label{massreach}
\end{table}

The mass reach estimated to be achieved through the single production process at $\sqrt{s}=1$ TeV with an integrated luminosity of 300  fb$^{-1}$ for selected final states (arising through single production channel) is given in Table~\ref{massreach-singleprod}. With the two selected channels of $2j+e^+e^-$ and $2b+e^+e^-$ arising from $\Sigma^\pm e^\mp$ production, and the final state $2b+E\!\!\!\!\! \slash$ arising from $\Sigma^0\nu$ could probe the model with $M_\Sigma$ close to 1 TeV, assuming $V_e=0.05$.

\begin{table}[h!]
\small
\begin{tabular}{|c||c||c||c||c||c||c||c||c||c|}
\hline \hline
Final State &$S$&$B$& $\sigma(\Sigma^+e^-)$&$M_\Sigma$&Final state & $S$&$B$& $\sigma(\Sigma^0 \nu)$&$M_\Sigma$\\
&&&in fb&in GeV& & &&in fb&in GeV\\ \hline \hline
$2j+e^- e^+$& 37 & 110 &  2.5 & 956  & $b\bar{b}+MET$ & 40.4 &   140.3 & 6.3  & 982 \\ \cline{1-5}
$b\bar{b}+e^- e^+$& 9 & 0 &  0.92 &  978 & &  & &  &  \\
\hline \hline
\end{tabular}
\caption{ The Mass reach at 1 TeV with integrated luminosity of 100 fb$^{-1}$ from selected channels of single production of charged and neutral  fermions, giving $3\sigma$ sensitivity. The corresponding production cross sections $\sigma(\Sigma^\pm e^\mp)$, $\sigma(\Sigma^0 \nu)$ and the number of signal ($S$) and background ($B$) events after the selection cuts are also given.}
\label{massreach-singleprod}
\end{table}

\section{Summary and Conclusions}
\label{conclusion}
Type III seesaw mechanism proposed to generate tiny neutrino mass provides an example of beyond the SM scenario with heavy leptons. 
We study the direct production of heavy leptons at high energy $e^+ e^-$ collider through  possible final states arising from their subsequent decays. Presently, direct searches at the LHC limits the masses of such heavy fermions in the range of 500 GeV or above. While the LHC is capable of discovering the presence of heavy leptons in the TeV mass range, it is hard to probe the details of the couplings involving mixing with the SM leptons.
 On the other hand, high energy $e^+ e^-$ colliders like the ILC  or CLIC with  electrons in the initial state  are suitable for this purpose, where the production process itself is sensitive to the mixing. Investigating the single and pair production of the neutral as well as charged leptons at $e^+e^-$ collider of centre of mass energies of 1 TeV and 2 TeV, respectively, this work performed detailed detector-level analyses to identify interesting final states, and the achievable significance for selected parameter choices.
 
Considering the single production $2b+e^+e^-$ final state is found to be the most promising, with the suitably chosen selection criteria completely eliminating the SM background. Among the other channels, $2j+e^-+E\!\!\!\!\! \slash$, $2b+E\!\!\!\!\! \slash$ and $e^+e^- +E\!\!\!\!\!\slash$ could be probed at $5\sigma$ significance (assuming only statistical uncertainty) with 2 to 4 fb$^{-1}$ luminosity, whereas the $2j+e^+e^-$ channel require about 14 fb$^{-1}$ luminosity. These conclusions assume a triplet lepton mass of $M_\Sigma=500$ GeV. Extrapolating this result to higher values of $M_\Sigma$, we found that a 1 TeV $e^+e^-$ collider with 300 fb$^{-1}$ luminosity could probe the mass very close to the kinematic limit of about 950 to 980 GeV, assuming $V_e=0.05$. Similarly, suitable selection of kinematic regions eliminate the SM background to the final states of 
$4j+e^+e^-$ and $4b+e^+e^-$ arising from the pair production of charged leptons, at the same time retaining sufficient number of signal events so as to have $5\sigma$ significance at 50 and 130 fb$^{-1}$ luminosities, respectively.  The corresponding luminosity in the case of muonic final states, $4j+\mu^+\mu^-$ and $4b+\mu^+\mu^-$, enabled by the mixing scenario of $V_e=0,~V_\mu=0.05$, is about 50 fb$^{-1}$. Other promising channels of $4j+\ell+E\!\!\!\!\!\slash$ and $2j+2\ell+E\!\!\!\!\!\slash$ require about 100 and 180  fb$^{-1}$ luminosities, respectively in the case of $\ell = e$, and about 50 and 180 fb$^{-1}$, in the case of $\ell = \mu$. The channel $2j+\mu^\pm+E\!\!\!\!\!\slash$ spares much better in the case of $V_\mu \ne 0$, requiring only about 70 fb$^{-1}$ luminosity for $5\sigma$ significance, whereas in the case of $V_e\ne 0$, the channel $2j+e^\pm+E\!\!\!\!\!\slash$ requires luminosity close to 500 fb$^{-1}$ to achieve the same significance.
The purely hadronic final state $4j+E\!\!\!\!\!\slash$ require 250 and 130 fb$^{-1}$ luminosities for the two cases of mixing scenarios with $V_e \ne 0$ and $V_\mu \ne 0$, respectively. Coming to the reach of $M_\Sigma$ through the pair production, the two channels, $4b+e^+e^-$ and $4j+e^+e^-$ could probe beyond 900 GeV with 300 fb$^{-1}$ luminosity, whereas all the channels with muonic final states could probe very close to the kinematic reach, going above 990 GeV.   

The study has clearly demonstrated the potential of high energy $e^+e^-$ collider to probe the presence of heavy leptons, and the details of their couplings with the SM particles,  thus supporting the case for such leptonic collider even with successful running of the LHC. Considering the nature of the process, with the presence of $t$-channel production in some of the mixing scenarios, we anticipate that beam polarisation could be utilised to enhance the sensitivity. Study of the effect of beam polarisation, as well as the detailed analysis to understand the reach on coupling is deferred to a future publication. 

\begin{acknowledgments}
This work is partly supported by the BRNS, DAE, Govt. of India project (2010/37P/49/BRNS/1446), and the SERB, DST of India project (EMR/2015/000333). The authors are thankful to Dr. Sumit K. Garg for useful discussions and involvement in the initial stage of the work.
\end{acknowledgments}

\section{Appendix}
\noindent
{\large \bf A1: The Lagrangian}
\label{appendix}

The Lagrangian of the Type III seesaw model in the mass basis is given below,
\begin{equation}
\mathcal{L} =  \mathcal{L}_{Kin} + \mathcal{L}_{CC} + \mathcal{L}_{NC}^{\ell} + \mathcal{L}_{NC}^{\nu} + \mathcal{L}_{H}^{\ell} + \mathcal{L}_{H}^{\nu} + \mathcal{L}_{\eta}^{\ell} + \mathcal{L}_{\eta}^{\nu} + \mathcal{L}_{\phi^{-}},
\end{equation}
where $\mathcal{L}_{Kin}$ is the kinetic part and 
\begin{eqnarray}
\mathcal{L}_{CC} & = & \frac{g}{\sqrt{2}} \begin{pmatrix} \bar{\ell} &  \overline{\Psi}\end{pmatrix} \gamma^{\mu}~W^{-}_{\mu} \left(P_{L}~g_{L}^{CC} + P_{R}~g_{R}^{CC}\sqrt{2} \right) \begin{pmatrix} \nu \\ \Sigma \end{pmatrix}  +h.c \\
\mathcal{L}^{\ell}_{NC} & = & \frac{g}{\cos\theta_{W}} \begin{pmatrix} \overline{\ell} & \overline{\Psi}\end{pmatrix} \gamma^{\mu}Z_{\mu} \left(P_{L}~g_{L}^{NC} + P_{R}~g_{R}^{NC} \right) \begin{pmatrix} \ell \\ \Psi \end{pmatrix}\\
\mathcal{L}^{\nu}_{NC} & = & \frac{g}{2\cos{\theta}_{W}}\begin{pmatrix}
\bar{\nu} & \overline{\Sigma}^{0c}\end{pmatrix} \gamma^{\mu}Z_{\mu}\left(P_{L}~g_{\nu}^{NC}\right)
\begin{pmatrix}
\nu_{L} \\ \Sigma^{0c}
\end{pmatrix} \\
\mathcal{L}^{\ell}_{H} &=& - \begin{pmatrix}
\overline{\ell} & \overline{\Psi}\end{pmatrix}H\left(P_{L}~g_{L}^{H\ell} + P_{R}~g_{R}^{H\ell} \right) \begin{pmatrix}
\ell \\ \Psi \end{pmatrix}\\
\mathcal{L}_{H}^{\nu} & = & - \begin{pmatrix}\overline{\nu} & \overline{\Sigma}^{0}
\end{pmatrix}\frac{H}{\sqrt{2}}\left(P_{L}~g_{L}^{H\nu} + P_{R}~g_{R}^{H\nu}\right)
\begin{pmatrix}
\nu \\ \Sigma^0 \end{pmatrix}\\
\mathcal{L}_{\eta}^{\ell} & = & - \begin{pmatrix}
\overline{\ell} & \Psi \end{pmatrix} i\eta \left( P_{L}~g_{L}^{\eta\ell} + P_{R}~g_{R}^{\eta\ell}\right)\begin{pmatrix} \ell \\ \Psi \end{pmatrix}\\
\mathcal{L}_{\eta}^{\nu} & = & - \begin{pmatrix}
\overline{\nu} & \overline{\Sigma}^{0} \end{pmatrix}\frac{i\eta}{\sqrt{2}}\left( P_{L}~g_{L}^{\eta\nu} + P_{R}~g_{R}^{\eta\nu}\right)\begin{pmatrix}\nu \\\Sigma^{0} \end{pmatrix}\\
\mathcal{L}_{\phi^{-}} & = & - \begin{pmatrix}
\overline{\ell} & \overline{\Psi} \end{pmatrix}\phi^{-} \left(P_{L}~g_{L}^{\phi^{-}} + P_{R}~g_{R}^{\phi^{-}} \right) \begin{pmatrix}
\nu \\ \Sigma^{0} \end{pmatrix}+h.c.
\end{eqnarray}
where, the left and right projection operates are denoted by, $P_{L,R} = \frac{1}{2} (1 \mp \gamma_5)$. The couplings, $g_i$ are explicitly given below in terms of the other parameters of the original Lagrangian.
Here, the fields and the couplings, $g_i$ are given in  block matrix form, with 
\begin{center}
$ \begin{pmatrix} \ell \\ \Psi \end{pmatrix}$ $\equiv$ $\begin{pmatrix}   e \\ \mu \\ \tau \\ \Psi \end{pmatrix}$ 
 and 
 $ \begin{pmatrix} \nu \\ \Sigma \end{pmatrix}$ $\equiv$ $\begin{pmatrix}   \nu_e \\ \nu_\mu \\ \nu_\tau \\ \Sigma^0 \end{pmatrix}$
 \end{center}

\begin{eqnarray}
g_{L}^{CC} & = & \begin{pmatrix}
\left(1+ \frac{\epsilon}{2}\right)U_{PMNS} & - \frac{\upsilon}{\sqrt{2} M_{\Sigma} }Y^{\dagger}_\Sigma \\ 0 & \sqrt{2}\left(1- \frac{\epsilon^{\prime}}{2}\right)\end{pmatrix}
\label{eq:couplings_b} \\
g_{R}^{CC} & = & \begin{pmatrix}
0 & -\frac{m_{\ell} \upsilon}{M_{\Sigma}^2}Y^{\dagger}_\Sigma \\   -\frac{\upsilon}{\sqrt{2} M_{\Sigma}}Y^{*}_{\Sigma}U^{*}_{PMNS} & 1- \frac{\epsilon^{\prime\star}}{2}
\end{pmatrix}\\
g_{L}^{NC} & = & \begin{pmatrix}
\frac{1}{2}- \cos^{2}\theta_{W} - \epsilon & 
\frac{\upsilon}{2 M_{\Sigma}}Y^{\dagger}_{\Sigma} \\ \frac{\upsilon}{2 M_{\Sigma}}Y_{\Sigma} & \epsilon^{\prime}-\cos^{2}\theta_{W}
\end{pmatrix}\\
g_{R}^{NC} & = & \begin{pmatrix} 1- \cos^{2}\theta_{W} & \frac{m_{\ell} \upsilon}{M_{\Sigma}^2}Y^{\dagger}_{\Sigma} \\
\frac{m_{\ell}\upsilon}{M_{\Sigma}^{2}}Y_{\Sigma} & - \cos^{2}\theta_{W}
\end{pmatrix}\\
g_{\nu}^{NC} & = & \begin{pmatrix}
1-U^{\dagger}_{PMNS}~\epsilon~ U_{PMNS}& \frac{\upsilon}{ \sqrt{2}~M_{\Sigma}}U^{\dagger}_{PMNS}Y^{\dagger}_{\Sigma} \\
\frac{\upsilon}{\sqrt{2 }M_{\Sigma}}Y_{\Sigma}U_{PMNS} & \epsilon^{\prime}
\end{pmatrix}\\
g_{L}^{H\ell} & = & \begin{pmatrix}
\frac{m_{\ell}}{\upsilon}(1-3\epsilon) & \frac{m_{\ell}Y^{\dagger}_{\Sigma}}{M_{\Sigma}} \\
Y_{\Sigma}\left(1-\epsilon + \frac{m_{\ell}^{2}}{M_{\Sigma}^2}\right) & \frac{\upsilon}{M_{\Sigma}}Y_{\Sigma}Y_{\Sigma}^{\dagger} \end{pmatrix}\\
g_{R}^{H\ell} & = & (g_{L}^{H\ell})^{\dagger} \\
g_{L}^{H\nu} &=&  \begin{pmatrix}
-\frac{\sqrt{2}~m^d_\nu}{\upsilon}&  \frac{m_{\nu}}{M_{\Sigma}}U^{T}_{PMNS}Y^{\dagger}_{\Sigma} \\
Y_{\Sigma}(1- \frac{\epsilon}{2} - \frac{\epsilon^{\prime}}{2})U_{PMNS} & \frac{\upsilon}{\sqrt{2}M_{\Sigma}}Y_{\Sigma}Y^{\dagger}_{\Sigma}
\end{pmatrix} \\ 
g_{R}^{H\nu} & = & (g_{L}^{H\nu})^{\dagger} \\
g_{L}^{\eta\ell} & = & \begin{pmatrix}
-\frac{m_{\ell}}{\upsilon}(1+\epsilon) & -\frac{m_{\ell}}{M_{\Sigma}}Y_{\Sigma}^{\dagger} \\ Y_{\Sigma} (1-\epsilon-\frac{m_{\ell}^{2}}{M_{\Sigma}^{2}})& \frac{\upsilon}{M_{\Sigma}} Y^{\dagger}_{\Sigma}Y_{\Sigma}
\end{pmatrix}\\
g_{R}^{\eta\ell} & = & -\left( g_{L}^{\eta \ell}\right)^{\dagger}\\
g_{R}^{\eta\nu} &= & - \left( g_{L}^{\eta \nu} \right)^{\dagger}\\
g_{L}^{\eta\nu} &=& g_{L}^{H\nu}\\
g_{L}^{{\phi}^{-}} &= &\begin{pmatrix}
\sqrt{2}\frac{m_{\ell}}{\upsilon}(1-\frac{\epsilon}{2})U_{PMNS} & \frac{m_{\ell}}{M_{\Sigma}}Y^{\dagger}_{\Sigma} \\ \frac{\sqrt{2}m_{\ell}^{2}}{M_{\Sigma}^{2}} Y_{\Sigma}U_{PMNS} & 0
\end{pmatrix}\\
g_{R}^{\phi^{-}} &=& \begin{pmatrix}
-\frac{\sqrt{2} m_{\nu}^{d*}}{\upsilon} U_{PMNS} & Y^{\dagger}_{\Sigma}(1 - \epsilon  - \frac{\epsilon^{\prime \star}}{2} -  \frac{2m^{\star}_{\nu}}{M_{\Sigma}})\\ -\sqrt{2}Y^{*}_{\Sigma}(1-\frac{\epsilon^{\star}}{2})U_{PMNS}^{\star} & 2\left(-\frac{M_{\Sigma}}{\upsilon} \epsilon^{\prime} + \epsilon^{\prime} \frac{M_{\Sigma}}{\upsilon} \right)
\end{pmatrix}
\label{eq:couplings_f}
\end{eqnarray}
 
Here, $\upsilon \equiv \sqrt{2}\langle\phi^{0}\rangle $ is the vev of the doublet scalar field, $\epsilon =  \frac{\upsilon^2}{M_{\Sigma}^2} Y_{\Sigma}^{\dagger}Y_{\Sigma}$,  $\epsilon'=  \frac{\upsilon^2}{2 M_{\Sigma}^2} \sum\limits_{\ell}Y_{ \Sigma \ell}^2 $ and  $U_{PMNS}$ is the lepton mixing matrix. The Yukawa coupling matrix $Y_\Sigma = 	\left(Y_{\Sigma e}~~Y_{\Sigma \mu}~~Y_{\Sigma \tau}\right)$, where $Y_{\Sigma \ell}$ are the Yukawa couplings appearing in Eq.~\ref{eq:lagrangian}. The mixing of $\Sigma$ with the SM leptons are denoted  by $V_\ell = \frac{\upsilon}{\sqrt{2}M_{\Sigma}} Y_{\Sigma\ell}$, where $\ell = e, \mu, \tau$.

\vspace{1cm}
\noindent
{\large \bf{A2: Cross sections for single and pair productions of fermions}  }

Expressions of the invariant amplitudes for the pair and single production of charged and neutral fermions are given below, with the general expression of cross-section given by,
\label{Apendix-csection}
\begin{center}
\Large{$\frac{d\sigma}{dt} =  {(4\pi \alpha)}^2\frac{\vert M \vert^2}{16 \pi s^2}$}.
\end{center}
\noindent
\textbf{1. Process $e^{+}e^{-} \rightarrow \Sigma^{-}\Sigma^{+}$}\\
The invariant amplitude for the pair production of charged fermion can be written as ,\\
\begin{eqnarray}
\vert M \vert^2 &=&  \frac{|M_{\gamma}|^2}{s^2} + \frac{1}{\cos^4\theta_{W} \sin^4 \theta_{W}} \left( \frac{|M_t|^2}{(t - m_{Z}^2)^2} + \frac{|M_s|^2}{(s-m_{Z}^2)^2} \right) +    \frac{1}{\cos^4\theta_{W} \sin^4 \theta_{W}}\left( \frac{M_{int}^{szt}} {(s- m_{Z}^2)(t - m_{Z}^2)} \right) \nonumber \\
&&
+  \frac{1}{\cos^2 \theta_{W} \sin^2 \theta_{W}} \left( \frac{M_{int}^{\gamma z}} {s(s - m_{Z}^2)} + \frac{M_{int}^{\gamma t}}{s(t - m_{Z}^2)}\right),
\end{eqnarray}
 where
\begin{eqnarray}
 |M_{t}|^2 &=& {\left(g_{L14}^{NC}\right)}^4  \left( 64 \left({(s +t)}^2 + m_{\Sigma}^{2} (  m_{\Sigma}^{2} - 2 s -2t ) \right) + \frac{s^2}{m_{Z}^2}\left(4s + t^2 +  \frac{m_{\Sigma}^{4}}{m_{Z}^2}  - 2t \frac{m_{\Sigma}^{2}}{m_{Z}^2} \right)(1-\beta^2)^2 \right)  \nonumber  \\ [10pt]
|M_{sz}|^2  &=& 64 ~~g_{L44}^{NC}~ ~g_{R44}^{NC}~  \left({(g_{R11}^{NC})}^2  +  {(g_{L11}^{NC})}^2  \right) (m_{\Sigma}^4 - s ~m_{\Sigma}^2 -2tm_{\Sigma}^2 +t^2) \nonumber \\&& + 64 \left( (g_{R11}^{NC}~~ g_{R44}^{NC})^2 + (g_{L11}^{NC}~~ g_{L44}^{NC})^2\right) ( m_{\Sigma}^4 - s~ m_{\Sigma}^2 -2t~m_{\Sigma}^2  + (s + t )^2 )   \nonumber \\ [10pt]
|M_{s\gamma}|^2 &=& 8s (s +2t) + 16~( t -m_{\Sigma}^2)^2 \nonumber \\ [10pt]
M_{int}^{szt} &=&  \frac{g_{L11}^{NC}~~{(g_{L14}^{NC})}^2}{m_{Z}^2} \left(32~ g_{R44}^{NC}~ ~ {m_{\Sigma}^2} \left( ( t - m_{\Sigma}^2)^2 - 2s m_{Z}^2 \right) \right.  \nonumber \\&&  +~ g_{L44}^{NC} \left.   \left( {32 s} \left((2 s + 2t -1) - (s + t)^2 \right) -  \frac{ s^4}{m_{Z}^2}(1-\beta^2)^2 \right) \right)   \nonumber \\ [10pt]
M_{int}^{\gamma z}  &=& 32~ \left( g_{L11}^{NC} ~~ g_{L44}^{NC} +  g_{R11}^{NC} ~~g_{R44}^{NC} \right)  \left( s(s -  m_{\Sigma}^2 + 2t) + ( m_{\Sigma}^2 - t)^2 \right)\nonumber \\&& + 32~ \left( g_{L11}^{NC} ~~g_{R44}^{NC} + g_{L44}^{NC}~~ g_{R11}^{NC} \right) \left(  s {m_{\Sigma}}^2 + (m_{\Sigma}^2 -  t)^2 \right) \nonumber \\ [10pt]
M_{int}^{\gamma t} &=& {(g_{L14}^{NC})}^2  \left( 32 m_{\Sigma}^2~ (s + 2t -m_{\Sigma}^2) - 32 (s + t)^2- \frac{s^3}{m_{Z}^2}(1 - \beta^2)^2 - \frac{16 m_{\Sigma}^2}{m_{Z}^2}~~(m_{\Sigma}^2 - t)^2 \right).\nonumber \\  \nonumber 
\end{eqnarray}

Here, $M_t$ is the invariant amplitude for the $t$-channel process, $M_{sz}$ and $M_{s\gamma}$  are invariant amplitudes for $s$-channel  processes with $Z$ boson and photon propagators, respectively (see Fig.~\ref{fig:fd-pair}). $M_{int}^{szt}$ gives the interference of $t$-channel with $s$-channel with  $Z$ boson propagator.   $M_{int}^{\gamma z}$ and $M_{int}^{\gamma t}$ give the interference of $s$-channel having photon propagator with the $s$-channel having $Z$ boson, and the 
$t$-channel processes, respectively. The factor $g_{ij}$'s are the corresponding elements of the coupling matrix given in Eq.~\ref{eq:couplings_b}-\ref{eq:couplings_f}.
\item  \textbf{2. Process $e^{+}e^{-} \rightarrow e^{-}\Sigma^{+}$}\\
The invariant amplitude for the single production of charged fermion can be written as ,\\ 
\begin{eqnarray}
\vert M \vert^2  &=&  \frac{1}{\cos^4\theta_{W} \sin^4 \theta_{W}} \left( \frac{|M_t|^2}{(t - m_{Z}^2)^2} + \frac{|M_s|^2}{(s-m_{Z}^2)^2}  + \frac{M_{int}^{ts}} {(t - m_{Z}^2) (s-m_{Z}^2)} \right)
\end{eqnarray}

 where
\begin{eqnarray}
|M_{t}|^2 &=& 
 \left({(g_{R11}^{NC}~~g_{R14}^{NC})}^2 + {(g_{L11}^{NC}~~g_{L14}^{NC})}^2 \right) \left( \beta^2~(64s^4 -32ts) + \beta^4~(16t^2 -32ts) \right) +\nonumber \\&&
  64~s^2 \beta^2 \left (g_{L14}^{NC}~{(g_{R11}^{NC})}^2 ~g_{R14}^{NC} + g_{L14}^{NC} ~{(g_{L11}^{NC})}^2 ~g_{R14}^{NC}  \right)   \nonumber \\ [10pt]
  |M_{t}|^2 &=& 
 \left({(g_{R11}^{NC}~~g_{R14}^{NC})}^2 + {(g_{L11}^{NC}~~g_{L14}^{NC})}^2 \right)\left( \beta^2~(64s^4 -32ts) + \beta^4~(16t^2 -32ts) \right) +\nonumber \\&&
  64~s^2 \beta^2 \left (g_{L14}^{NC}~{(g_{R11}^{NC})}^2 ~g_{R14}^{NC} + g_{L14}^{NC} ~{(g_{L11}^{NC})}^2 ~g_{R14}^{NC}  \right)   \nonumber \\ [10pt]
 |M_{s}|^2 &=& 
 \left({(g_{R11}^{NC} ~g_{R14}^{NC})}^2 + {(g_{L11}^{NC} ~g_{L14}^{NC})}^2  \right) \left(\beta^2~(64~s^2 -32 t~s) + \beta^4 ~(16~ t^2 -32~t~s)\right) + \nonumber \\&&
  \left(g_{L14}^{NC} ~~{(g_{R11}^{NC})}^2 ~~ g_{R14}^{NC} + {(g_{L14}^{NC}~~ g_{R11}^{NC})}^2  \right)  \left( 32~ts \beta^2  - \beta^4(16t^2 -32ts)\right)  \nonumber \\ [10pt]
M_{int}^{t s}  &=&   \left({(g_{R11}^{NC} ~~g_{R14}^{NC})}^2 + {(g_{L11}^{NC} ~~g_{L14}^{NC})}^2  \right)\left( (64~t~s - 128 s^2) \beta^2 + (64 ~t~s - 32~ t^2)\beta^4)\right) \nonumber \\ \nonumber 
\end{eqnarray}
Here again, $M_t$,  $M_s$ are invariant amplitudes for the t-channel, s-channel  with Z boson propagator and $M_{int}^{t s}$  is the interference term involving  t- and s-channel with Z boson propagator shown in Fig \ref{fig:fd-single}.
\item \textbf{3. Process $e^{+}e^{-} \rightarrow \nu_{\ell}\Sigma^{0}$}\\
The invariant amplitude for the single production of neutral fermion can be written as ,\\ 
\begin{eqnarray}
 \vert M \vert^2 & = &  \frac{M_{\gamma}|^2}{s^2} + \frac{1}{4 \sin^4 \theta_{W}} \left( \frac{|M_t|^2}{(t - m_{W}^2)^2} \right) + \frac{1}{4 \cos^4\theta_{W} \sin^4 \theta_{W}} \left( \frac{|M_s|^2}{(s-m_{Z}^2)^2} \right)  +  \frac{1}{2 \sin^2 \theta_{W}} \left( \frac{M_{int}^{gtW}}{s(t - m_{W}^2)} \right) + \nonumber \\ &&   \frac{1}{4 \cos^2\theta_{W} \sin^4 \theta_{W}} \left( \frac{M_{int}^{ztW}} {(t - m_{W}^2)(s-m_{Z}^2)} \right) +  +   \frac{1}{2 \cos^2\theta_{W} \sin^2 \theta_{W}} \left( \frac{M_{int}^{gz}}{s(s-m_{Z}^2)} \right),
\end{eqnarray}
 where

\begin{eqnarray}
|M_t|^2 &=& 
 \left({(g_{R1\ell}^{CC} ~~g_{R14}^{CC})}^2 + {(g_{L1\ell}^{CC}~~ g_{L14}^{CC})}^2  \right) \left( \beta^2~(64~s^2-32~t~s ) + \beta^4 ~(16~t^2 - 32~t~s) \right) + \nonumber \\&&
 64 ~s^2\beta^2 ~\left({(g_{R1\ell}^{CC} ~~g_{L14}^{CC})}^2 + {(g_{L1\ell}^{CC} ~~g_{R14}^{CC})}^2  \right)  \nonumber \\
  |M_{s}|^2 &=&  16~t~ \beta^4~ {(g_{\nu 14}^{NC})}^2 (t - 2 s) \left({(g_{R11}^{NC})}^2 + {(g_{L11}^{NC})}^2  \right)+ 32 s~\beta^2 ~{(g_{\nu 14}^{NC})}^2 (2s - t) \left({(g_{L11}^{NC})}^2 + 32 ts {(g_{R11}^{NC})}^2 \right)
  \nonumber \\ \nonumber 
 \end{eqnarray}
 \begin{eqnarray}
 |M_\gamma |^2 &=& 4\beta^2 (s^2 + t\beta^2(t - 2s) ) \nonumber \\ [10pt]
M_{int}^{ztW} &=& 
32~  \beta^2  (2s - \beta^2~t) (  t - 2s) ~\left(g_{L1\ell}^{CC} ~ g_{L14}^{CC} ~ g_{L11}^{NC} ~ g_{\nu 14}^{NC} \right)   \nonumber \\ [10pt]
M_{int}^{gtW} &=& 
8~\beta^2  (t - 2 s ) (2s - t\beta^2)   \left( g_{R1\ell}^{CC}~ ~g_{R14}^{CC} +  g_{L1\ell}^{CC} ~~g_{L14}^{CC}  \right)  \nonumber  \\ [10pt]
M_{int}^{gz} &=&
 \beta^4~(g_{\nu 14}^{NC})~(8t^2 - 16 ~t~s)\left(g_{R11}^{NC}+ g_{L11}^{NC}\right) +    \beta^2 ~ g_{\nu 14}^{NC} \left( 64~t~s~(g_{R11}^{NC} - g_{L11}^{NC}) + 32~ s^2 ~g_{L11}^{NC}   \right) \nonumber \\ \nonumber 
\end{eqnarray}
Here, $M_t$,  $M_s$, $M_\gamma$ are the  invariant amplitudes with the propagator of $W$ boson ($t$-channel), $Z$ boson and photon, respectively . $M_{int}^{ztW}$ is the invariant amplitude of interference terms of the s-channel with  $Z$ boson propagator and  the t-channel with $W$ boson propagator. $M_{int}^{gtW}$ is the invariant amplitude of interference term with  s-channel containing  photon propagator  and the t-channel containing  $W$ boson propagator. $M_{int}^{gz}$ is the invariant amplitude of interference term having s-channel with $Z$ boson  and photon propagator ( see Fig \ref{fig:fd-single}).

\end{document}